\documentclass{aa}

\usepackage[utf8]{inputenc} 
\usepackage[varg]{txfonts}
\usepackage{amssymb}
\usepackage{epsfig}
\usepackage{graphics}
\usepackage{amsmath}
\usepackage{color}
\usepackage{natbib}
\usepackage{hyperref}
\usepackage{gensymb}

\usepackage{bm}
\usepackage{mathtools}
\usepackage{graphicx}
\usepackage{lipsum}

\usepackage{mathtools,tikz,caption}
\DeclareRobustCommand\sampleline[1]{%
  \tikz\draw[#1][line width=0.35mm] (0,0) (0,\the\dimexpr\fontdimen22\textfont2\relax)
  -- (2.2em,\the\dimexpr\fontdimen22\textfont2\relax);%
}
\usepackage[normalem]{ulem}
\usepackage{physics}

\newcommand{\be}{\begin{equation}}
\newcommand{\ee}{\end{equation}}
\newcommand{\beq}{\begin{eqnarray}}
\newcommand{\eeq}{\end{eqnarray}}

\newcommand{\red}[1]{\textcolor{red}{#1}}
\newcommand{\blue}[1]{\textcolor{blue}{#1}}

\newcommand{\orange}[1]{\textcolor{orange}{#1}}
\newcommand{\diff}{\mathrm{d}}

\newcommand{\source}{\object{SAX J1808.4$-$3658}}

\newcommand{\msun}{{M}_{\sun}}
\newcommand{\req}{R_{\mathrm{e}}}

\bibpunct{(}{)}{;}{a}{}{,} 

\DeclareUnicodeCharacter{00A0}{ }

\makeatletter
\def\fvec#1{\underline{\sbox\tw@{$#1$}\dp\tw@\z@\box\tw@}}
\makeatother

\begin{document}

\title{Neutron star parameter constraints for accretion-powered millisecond pulsars from the simulated IXPE data}

\titlerunning{Neutron star parameter constraints from IXPE data}

\author{Tuomo~Salmi\inst{1}
\and Vladislav Loktev\inst{1,2}
\and Karri Korsman\inst{1}
\and Luca Baldini\inst{3}
\and Sergey S. Tsygankov\inst{1,2}
\and  Juri~Poutanen\inst{1,2,4}}

\institute{Tuorla Observatory, Department of Physics and Astronomy, 20014 University of Turku, Finland\\ \email{thjsal@utu.fi} 
   \and Space Research Institute of the Russian Academy of Sciences, Profsoyuznaya str. 84/32, 117997 Moscow, Russia 
   \and Universit\`a di Pisa and Istituto Nazionale di Fisica Nucleare, Sezione di Pisa, Largo Bruno Pontecorvo, 3, Pisa 56127, Italy
   \and Nordita, KTH Royal Institute of Technology and Stockholm University, Roslagstullsbacken 23, SE-10691 Stockholm, Sweden
}

\date{Received 18 September 2020 / Accepted 22 November 2020}

\abstract{
We have simulated the X-ray polarization data that can be obtained with the Imaging X-ray Polarimetry Explorer, when observing accretion-powered millisecond pulsars.
We estimated the necessary exposure times for \source\ in order to obtain different accuracy in the measured time-dependent Stokes profiles integrated over all energy channels. 
We found that the measured relative errors strongly depend on the relative configuration of the observer and the emitting hotspot. 
The improvement in the minimum relative error in Stokes $Q$ and $U$ parameters as a function of observing time $t$ scales as $1/\sqrt{t}$, and it spans the range from 30--90\% with a 200 ks exposure time to 20--60\% with a 500~ks exposure time (in the case of data binned in 19 phase bins). 
The simulated data were also used to predict how accurate measurements of the geometrical parameters of the neutron star can be made when modelling only $Q$ and $U$ parameters, but not the flux.
We found that the observer inclination and the hotspot co-latitude could be determined with better than 10\degr\  accuracy for most of the cases we considered. 
In addition, we show that the position of a secondary hotspot can also be constrained when the spot is not obscured by an accretion disc.
These measurements can be used to further constrain the neutron star mass and radius when combined with modelling of the X-ray pulse profile.
}

\keywords{polarization -- stars: neutron  -- stars: atmospheres -- methods: numerical -- X-rays: binaries}

\maketitle

\section{Introduction}\label{sec:intro}

Accretion-powered millisecond pulsars (AMPs) are rapidly rotating neutron stars (NSs) located in low-mass X-ray binaries. 
These weakly magnetised NSs have been spun-up by the accretion torques acting on the NS during the process of gas accretion from its companion \citep{RS82,ACR82}. 
The accreting matter falls on the NS surface, creating hotspots at the magnetic poles that emit X-ray radiation. 
Rotation of the NS produces pulsation and the pulse shape contains information about the parameters of the NS such as its mass and radius  \citep[see e.g.][]{PG03,2016RvMP...88b1001W}, which can be used to constrain the equation of state (EOS) for extremely dense matter of inner parts of the NS \citep[for instance, see][]{lindblom1992,Lattimer12ARNPS,RRW_nicer19}. 

Previously, pulse profiles have been modelled in order to constrain the NS parameters \citep[see e.g.][]{PFC83,ML98,PG03,PB06,MLC07,lomiller13,SNP18,BLM_nicer19,MLD_nicer19,RWB_nicer19}. 
Usually, this approach has a problem to uniquely determine the pulsar geometry, particularly the observer inclination and the co-latitude of the emitting spot, which are highly degenerate with the NS mass and radius. 
However, the phase dependence of the polarization angle (PA), and therefore Stokes parameters $Q$ and $U$, is a powerful tool to constrain the NS geometry \citep{VP04,Poutanen10}.
The transformation of Stokes parameters from the NS surface to the observer frame was studied by \citet{VP04} and  \citet{poutanen20} for the case of a rapidly rotating spherical NS, and also for the case of an oblate star by \citet{LSNP20}.

The radiation escaping the NS surface in AMPs is expected to be significantly polarized, as it is scattered by the hot electrons in the accretion shock above the NS surface. 
In rotation-powered millisecond pulsars (RMPs), on the other hand, the dominant thermal radiation is too soft and likely weakly polarized; only a possible tail \citep{Salmi20} above a few keV may have a significant polarization, but it is too dim to be detected by the upcoming X-ray polarization instruments.   
The polarization degree (PD) produced by electron scattering strongly depends on the scattering angle and electron temperature \citep[see e.g.][]{NP94a,Pou94,Pou94ApJS}.
A model for polarized radiation from AMPs, based on Comptonization in an optically thin NS atmosphere, but in Thomson scattering approximation, was introduced in \citet{VP04} \citep[see also][]{ST85}. 
For a more accurate model, one should apply the formalism for Compton scattering in a hot slab \citep{PS96}; this is, however, computationally more expensive. 
We used a slightly modified version of the Thomson model (see Sect. \ref{sec:emission_model}) to make the first detailed simulations of the upcoming X-ray polarization observations and NS parameter constraints, which are expected to be measured soon. 
The results can be applied when designing the observational strategy of the Imaging X-ray Polarimeter Explorer \citep[IXPE;][]{IXPE} or other future X-ray polarimetric missions such as the enhanced X-ray Timing and Polarimetry mission \citep[eXTP;][]{Zhang19,dmatter_extp}.

For simplification, the case with one hotspot was applied in most of the calculations. 
This is justified because most of the AMPs expected to be observed by IXPE show simple sine-like profiles during the peak of their outbursts \citep{KM04,HPC08,LMC09,LMC11,SPB17}. 
This fact is naturally explained if only one spot is visible to the observer while the other is hidden by the accretion disc \citep{IP09,PIA09}. 
However, observations of \source\ show that the pulse profile starts to significantly deviate from the sine-wave at low fluxes, where the magnetosphere expands sufficiently so that we can see the secondary spot. 
Also, few other AMPs show signs of a secondary spot \citep[see e.g.][]{GC02,SPR18} 
and understanding the effects from multiple spots would also be useful for possible future X-ray polarization measurements of RMPs.  
Therefore, we also consider a few cases with two possibly non-antipodal spots, as suggested by the non-dipolar magnetic field configuration indicated by the observations of PSR~J0030+0451 by the Neutron star Interior Composition ExploreR \citep{BWH_nicer19,MLD_nicer19,RWB_nicer19}.

The remainder of this paper is structured as follows. 
In Sect. \ref{sec:pol_modelling}, we present the methods used to model polarized radiation from an AMP. 
In Sect. \ref{sec:ixpeobssim}, we explain how we simulated the data that could be observed by IXPE, and in Sect. \ref{sec:bayesian} we describe the Bayesian method used to obtain constraints on the NS model parameters.
The produced simulated data and the predicted NS parameter constraints are shown in Sect. \ref{sec:results}. 
A discussion and conclusions appear in Sects. \ref{sec:discussion} and \ref{sec:conclusions}, respectively.

\section{Methods}\label{sec:methods}

\subsection{Modelling polarization profiles}\label{sec:pol_modelling}

Our polarization modelling is mostly based on the polarization model introduced in \citet{VP04} and on the X-ray pulse shape modelling introduced in \citet{PB06}.
We used the `oblate Schwarzschild' approximation, which takes the deformed shape of the star into account in addition to the special and general relativistic corrections to the photon angles and trajectories \citep{MLC07,ML15,SNP18,Suleimanov20}. 
The shape of the NS was obtained from the model presented by \citet{AGM14}, which is suitable for the spin frequencies considered here (see \citealt{SPY2020} for a recent model for the most rapidly rotating stars). 
The description of polarized pulse formation has been slightly revised since \citet{VP04}, and it accounts for the effects of the oblate shape of the star on the observed PA \citep{LSNP20}. 
Otherwise, the Stokes parameters were computed in the frame comoving with the spot and then transformed to the observer frame as in \citet{VP04} and  \citet{poutanen20}. 
We note that the oblateness and relativistic effects are significant for stars spinning at a rate higher than 200~Hz, and assuming an incorrect shape may bias the constraints on NS geometry, as shown by \citet{LSNP20}.

We assumed that the observed photons originate in one or two spots at the NS surface, which we modelled as a slab of hot electrons above a blackbody emitting surface.  
The spectral energy distribution of the radiation was calculated as in \citet{SNP18} using the Comptonization model \textsc{simpl}  \citep{Steiner2009} from the {\sc xspec} package \citep{Arn96}, which is based on the solution of the non-relativistic Kompaneets equation \citep{ST80}.
The parameters of the model are the photon spectral index $\Gamma$ and a fraction $X_{\mathrm{sc}}$ of black-body photons scattered in the slab.
We used the version \textsc{simpl-2},  which takes both Compton up-scattered and down-scattered photons  into account.
The model converts the specified fraction of black-body photons (with temperature $T_{\mathrm{bb}}$) into a Comptonized power-law-like spectrum. 

The values of all the parameters in our fiducial model are shown in Table \ref{table:params}. 
Especially, the geometrical parameters are the observer inclination $i$ (angle between the spin axis and the line of sight), co-latitude of the primary spot $\theta$ (angle between the spin axis and the radius vector of the spot centre), pulsar rotation axis position angle $\chi$ (angle measured from the north counterclockwise to the projection of the rotation axis on the plane of the sky), and an arbitrary phase shift $\Delta \phi$. 
When simulating the data and calculating parameter constraints in Sect.~\ref{sec:results}, we also consider several sets of modified initial parameters, which are shown in Table \ref{table:all_models}. 
In the case of a two-spot model, we additionally considered the co-latitude of a secondary spot $\theta_{12}$ and the longitude difference between the two spots $\phi_{12}$.

\begin{table}
  \caption{Fiducial parameters of the synthetic data.} 
\label{table:params}
\centering
  \begin{tabular}[c]{ l  c } 
    \hline\hline 
     Parameter & Value\\ \hline
    \multicolumn{2}{c}{Neutron star parameters} \\ 
      Equatorial radius $\req$ & $12.0$ km  \\ 
      Mass $M$ & $1.4 ~\msun$  \\ 
      Spin frequency $\nu$ & 401 Hz \\
      Inclination $i$ & 60\degr \\ 
      Spot 1 co-latitude $\theta$ & 20\degr \\ 
      Spot angular radius $\rho$ & 1\degr  \\ 
  Pulsar rotation axis position angle $\chi$ & $0 \degree$ \\
        Phase shift $\Delta \phi$ & $0 \degree$ \\
      Spot 2 co-latitude $\theta_{2}$ & - \\
      Spot 2 longitude difference $\phi_{12}$ & - \\
      \hline
    \multicolumn{2}{c}{Emission model parameters} \\ 
      Electron gas temperature $T_{\mathrm{e}}$ & $50$ keV \\ 
      Thomson optical depth $\tau$ & 1 \\ 
      Seed photon temperature 
      $T_{\mathrm{bb}}$ & $1$ keV \\
      Scattered photon fraction $X_{\mathrm{sc}}$ & $0.6$  \\
      Photon spectral index $\Gamma$ & $1.8$  \\
    Maximum initial polarization $p_{\max}$ & 0 \\ 
    \hline
  \end{tabular}
  \end{table}

\begin{table}
  \caption{Parameters altered from the fiducial model for all of the computed models.}\label{table:all_models}
\centering
  \begin{tabular}[c]{l c c c c c c} 
    \hline\hline
     Model & $i$ & $\theta$ & $\theta_{2}$ & $\phi_{12}$ & $\rho$ & $p_{\max}$ \\ 
           & (deg) & (deg)  & (deg) & (deg) & (deg) & \\ \hline
           \multicolumn{7}{c}{One spot} \\
1\tablefootmark{a} \, \blue{\sampleline{solid}} & 60 & 20 & - &  - & 1 & 0 \\
2 \,\,\, \sampleline{dashed}  & 50 & 20 & - &  - & 1 & 0 \\
3 \,\,\, \sampleline{dash pattern=on 5pt off 1pt on 1pt off 1pt} & 70 & 20 & - &  -& 1 & 0 \\
4 \,\,\, \red{\sampleline{dashed}} & 60 & 10 & - &  - & 1 & 0 \\
5  \,\,\, \red{\sampleline{dash pattern=on 5pt off 1pt on 1pt off 1pt}} & 60 & 30 & - &  -& 1 & 0 \\
6 \,\,\, \orange{\sampleline{solid}} & 60 & 20 & - &  - & 30 & 0 \\
7 \,\,\, \blue{\sampleline{dash pattern=on 5pt off 1pt on 1pt off 1pt}} & 60 & 20 & - &  - & 1 & $0.1171$ \\
\hline
\multicolumn{7}{c}{Two spots} \\
8 & 60 & 20 & 160 & 180 & 1 & 0 \\
9 & 60 & 20 & 120 & 180 & 1 & 0 \\
\hline 
 \end{tabular}
\tablefoot{
The presented line styles correspond to those shown in Figs.~\ref{fig:data_errors_rel} and \ref{fig:data_errors_sigma} for one-spot models. 
We also note that the emission model parameters and the spot angular radii are the same for both spots in the two-spot models. 
\tablefoottext{a}{Fiducial set of parameters from Table \ref{table:params}. }
}
\end{table}

\subsubsection{Emission model}\label{sec:emission_model}
The dependence of the PD on the zenith angle (or its cosine $\mu$) and photon energy $E$ (in the spot frame) is still not well-known for the AMPs.
Here, we used a slightly modified version of the simple Thomson scattering model in a plane-parallel atmosphere presented in \citet{VP04}.
In the co-moving frame of the spot, the PD $P(E,\mu)$ as a function of photon energy $E$ and cosine of the zenith angle $\mu$ was obtained from the ratio of Stokes parameters $Q$ and $I$ (Stokes $U$ is zero due to the assumed azimuthal symmetry of the radiation) for photons scattered multiple times in a slab of Thomson optical depth $\tau = 1$ and electron gas temperature $T_{\mathrm{e}} = 50$~keV. 
The energy of the $n$-times scattered photon can be found from \citep{RL79}
\be \label{eq:e_nscat}
 \frac{E_n}{E_{0}} = A^n= \left( 1 + \frac{4kT_{\mathrm{e}}}{m_{\mathrm{e}}c^{2}} \right)^{n},
\ee
where $E_{0}$ is the energy of a seed photon and $A$ is the amplification factor for each scattering. 

We assumed the seed photons to have an energy distribution described by the Planck function of temperature $T_{\mathrm{bb}} = 1$~keV with the angular distribution given by the function $a_{\mathrm{s}}(\mu)$: 
\be \label{eq:emit_elsc}
I_{\mathrm{s}}(\mu,E) = a_{\mathrm{s}}(\mu)\ B_{E}(T_{\mathrm{bb}})\ . 
\ee 
We considered two cases: (1) the isotropic intensity with the constant angular function $a_{\mathrm{s}}(\mu)=1$ and (2) the angular distribution corresponding to the electron-scattering dominated semi-infinite atmosphere \citep{1947ApJ...105..435C,Cha60,sob49,Sob63}:  
\be \label{eq:amu_elsc}
a_{\mathrm{s}}(\mu) =0.421+0.868\,\mu.
\ee 
In the first case, we assumed unpolarized emission $P_{\mathrm{s}} (\mu)=0$. In the second case, we took \citep{VP04}
\beq \label{eq:pol_deg}
P_{\mathrm{s}} (\mu)& = & -  \frac{1-\mu}{1+3.582 \mu} \  p_{\max}, 
\eeq
where $p_{\max}=11.71$\%. 
For weakly magnetised NSs, the intensity of the radiation in the spot comoving frame is expected to be independent of azimuth. 
Therefore, the polarization of the radiation (i.e. dominant direction of electric field oscillations) is either in the meridional plane, containing the normal to the surface and the line-of-sight, or perpendicular to that plane. 
We define the polarization as positive in the first case and negative in the second.

After leaving the slab, the unscattered photons have the following angular dependence of intensity: 
\be  \label{eq:unscat_int}
a_{l,r}^{0}(\mu) = a_{\mathrm{s}}(\mu) 
\ \mbox{e}^{-\tau/\mu}\ \left( 1\pm P_{\mathrm{s}}(\mu)\right) ,
\ee
where indices l, r (as well as the $\pm$ sign) refer to the intensity of radiation polarized in the meridional plane and perpendicular to it, respectively. 
The angular dependence of the total intensity (also known as the beaming function) for escaping unscattered photons is $a^{0}(\mu) = a_{l}^{0}(\mu)+a_{r}^{0}(\mu)$. 

The energy dependence of the radiation escaping from the slab was accounted for by assuming that the dependence of intensity on $\mu$ and $E$ can be separated:
\be 
I_{l,r}(\mu,E)= a^{n}_{l,r}(\mu) \epsilon^{n}(E).
\ee
The angular dependencies of the intensities for remaining scattering orders $a^{n}_{l,r}(\mu)$ were calculated based on the intensities of the zeroth scattering order using the formulae presented in \citet{VP04}. 
The energy function was approximated with a diluted Planck function assuming that in each scattering, a photon increases its energy by a factor of $A$: 
\be \label{eq:epsilon_n}
\epsilon^{n+1}(E)= \epsilon^{n}(E/A),  
\ee
with $\epsilon^{0}(E)=B_E(T_{\mathrm{bb}})$. 
Such a relation implies that the number of photons 
$\int (\epsilon^{n}(E)/E )\diff E$ does not depend on $n$.
In reality, of course, the number of scattered photons decreases with $n$, but this was accounted for in the normalization of the angular functions $a^{n}_{\rm l,r}(\mu)$. 
Thus, the corresponding Stokes parameters in this approximate Thomson scattering model $I_{\mathrm{T}}$ and $Q_{\mathrm{T}}$ are given by: 
\be \label{eq:iq_thom_sc}
\left( 
\begin{array}{c}
I_{\mathrm{T}}^{n}(\mu,E) \\
Q_{\mathrm{T}}^{n}(\mu,E) \\
\end{array} \right) =
\left( \begin{array}{c}
a_{l}^{n}(\mu)+a_{r}^{n}(\mu) \\
a_{l}^{n}(\mu)-a_{r}^{n}(\mu) \\ 
\end{array} \right) \epsilon^{n}(E). 
\ee
For a given energy $E$, we obtained the Stokes parameters by summing Eq. \eqref{eq:iq_thom_sc} over $n$ as
\be \label{eq:sum_scat_orders}
I_{\mathrm{T}}(\mu,E) = \sum_{n=0}^{N} I_{\mathrm{T}}^{n}(\mu,E), \qquad 
Q_{\mathrm{T}}(\mu,E) = \sum_{n=0}^{N} Q_{\mathrm{T}}^{n}(\mu,E),
\ee
where we used $N=23$ for the highest number of scatterings. 
The final PD is then 
\be \label{eq:pd_final}
P(\mu,E) = Q_{\mathrm{T}}(\mu,E)/I_{\mathrm{T}}(\mu,E).
\ee 
We emphasise that these Stokes parameters were only used when evaluating the PD as a function of energy and the emission angle, and they do not provide the final energy dependence of the radiation.

\begin{figure*}
\centering
\includegraphics[width=8cm]{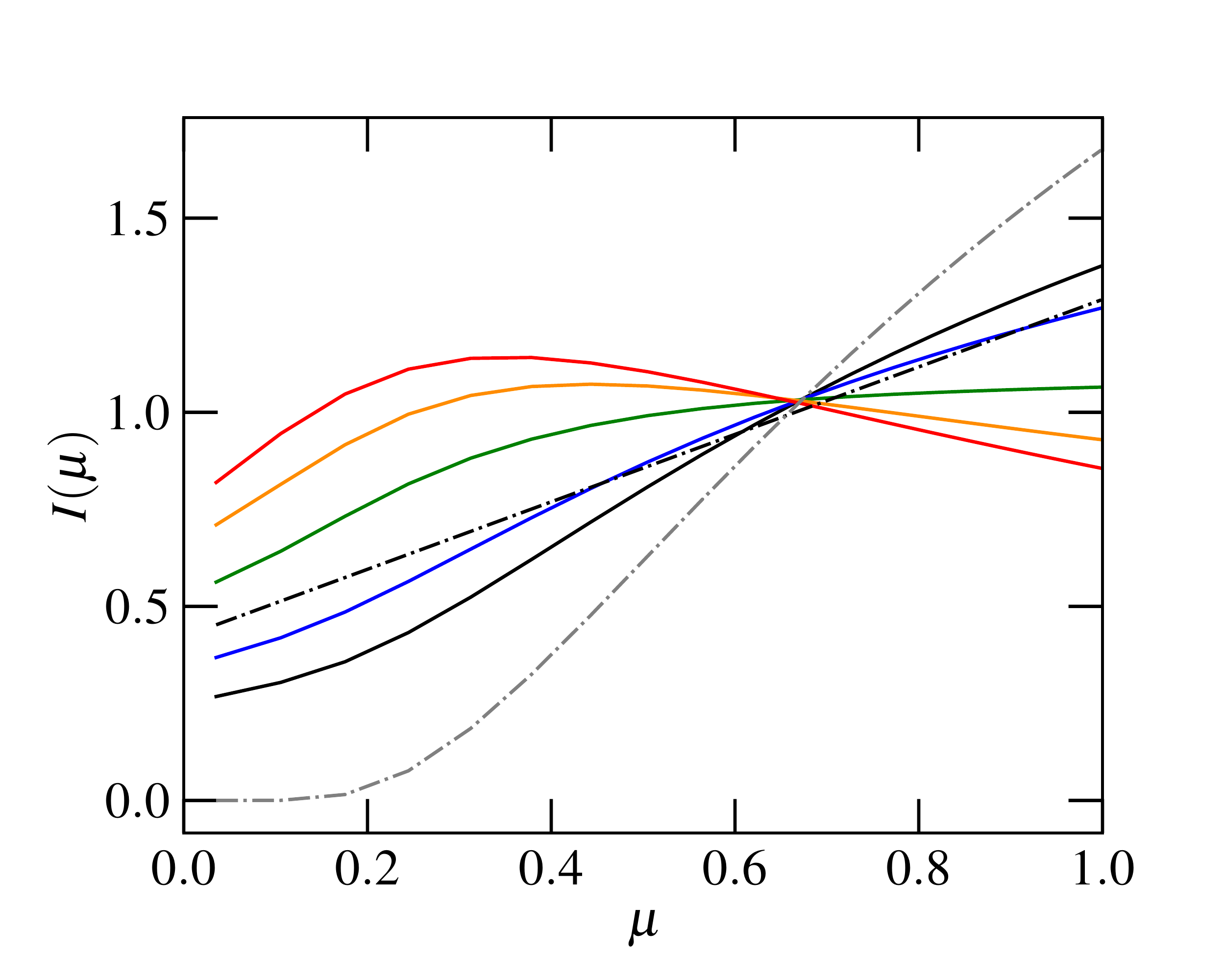} 
\includegraphics[width=8cm]{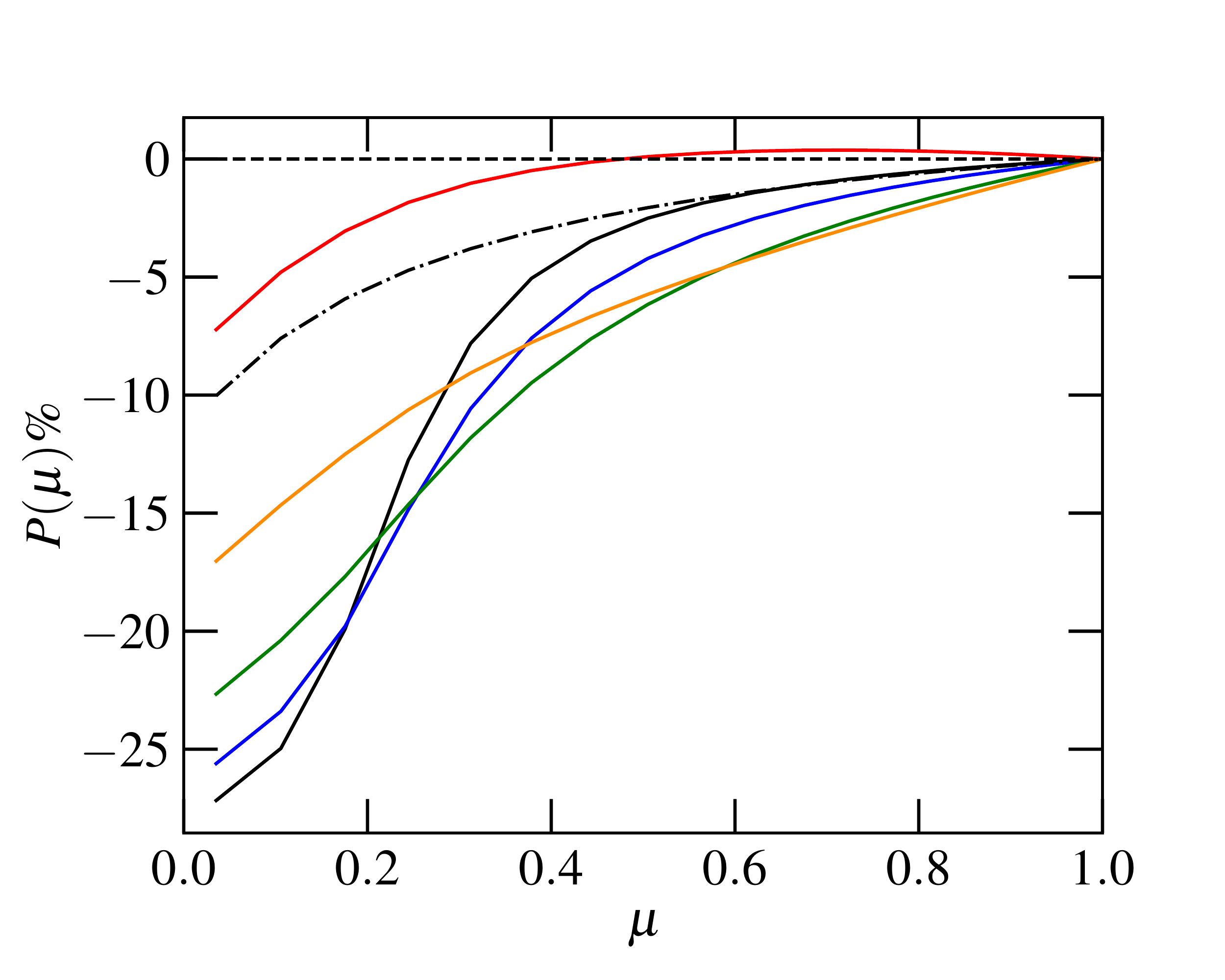}
\caption{
Angular dependence of emitted intensity (\textit{left panel}) and PD (\textit{right panel}). 
The angular dependence was normalized so that $\int_{0}^{1}\mu I(\mu) \diff \mu  = 1/2$. 
Black, blue, green, orange, and red solid lines show the model for photon energies of 2, 5, 8, 12, and 18 keV, respectively. 
The black dot-dashed curves show the intensity and polarization corresponding to the classical results of Chandrasekhar-Sobolev (see Eq.~\eqref{eq:pol_deg}), corresponding to the optically thick electron-scattering dominated atmosphere.
The grey dot-dashed curve in the \textit{left panel} shows the angular dependence of intensity for unscattered photons.
The model was calculated with the parameters shown in Table \ref{table:params} (e.g. $p_{\max}=0$). 
}
\label{fig:beaming}
\end{figure*}

\begin{figure*}
\centering
\includegraphics[width=8cm]{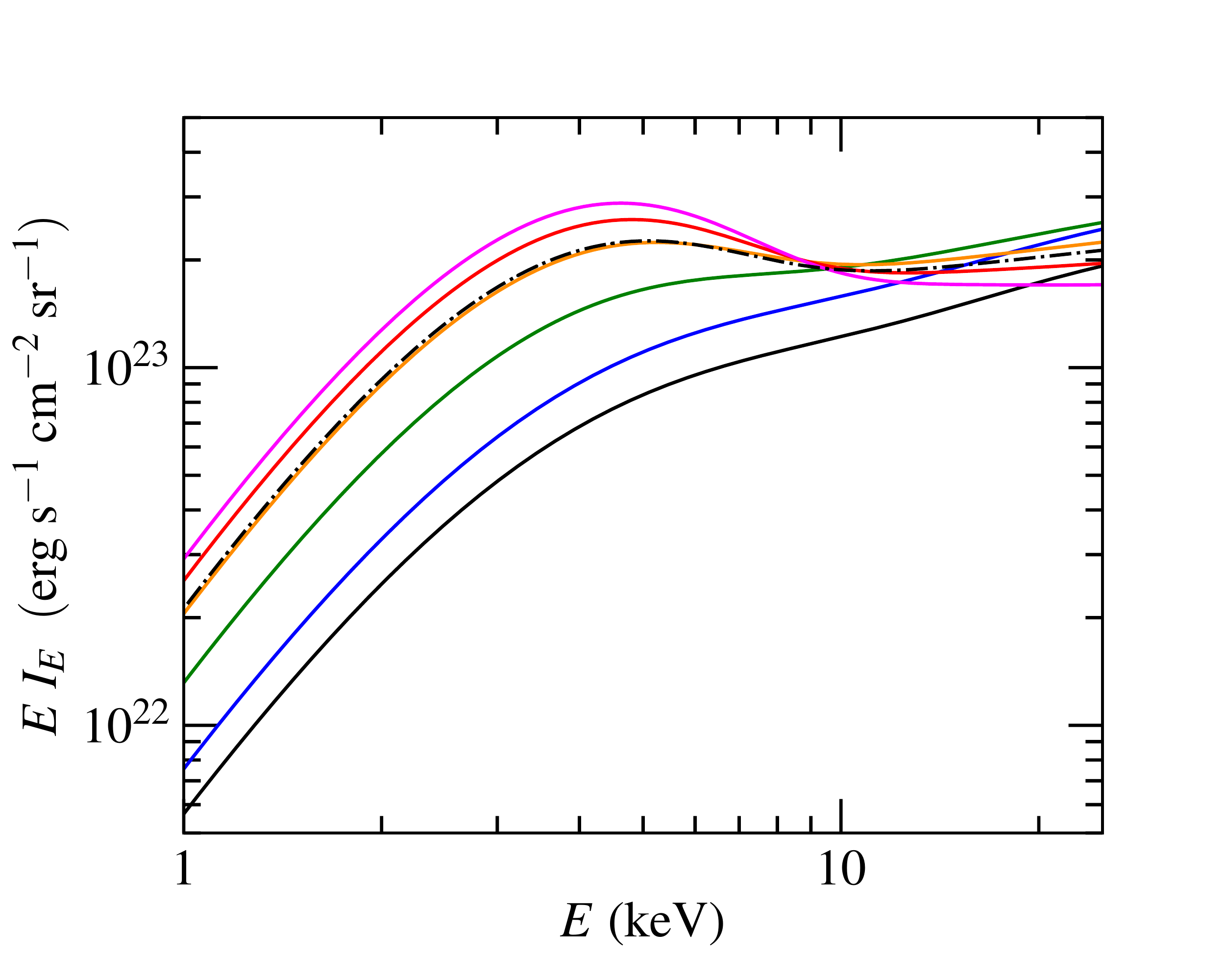} 
\includegraphics[width=8cm]{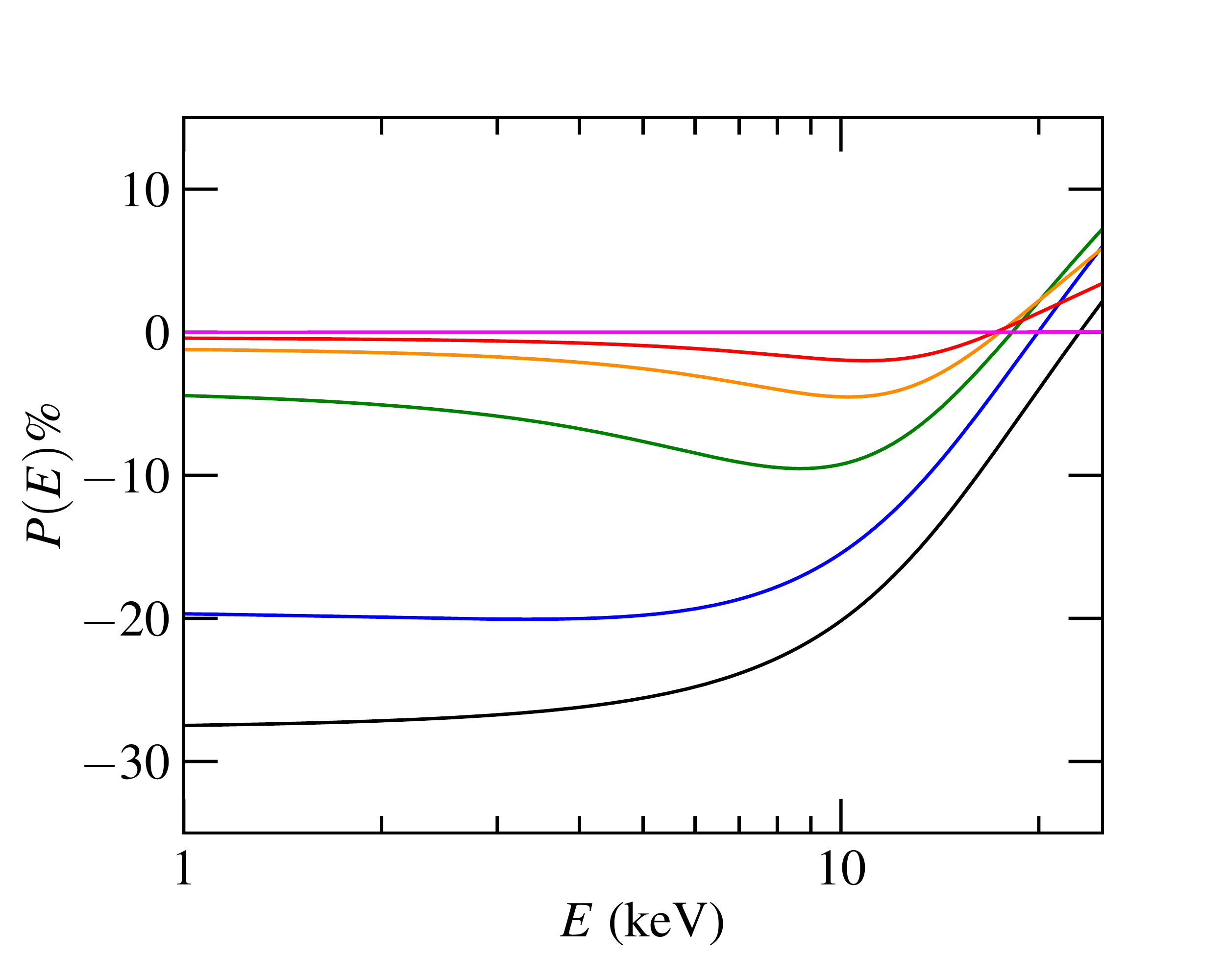}
\caption{
Energy dependence of emitted intensity (\textit{left panel}) and PD (\textit{right panel}). 
Black, blue, green, orange, red, and magenta solid lines show the model for $\mu = $ 0.0, 0.2, 0.4, 0.6, 0.8, and 1.0, respectively. 
The dot-dashed black curve (\textit{left}) shows the intensity corresponding to the angle-independent result of model \textsc{simpl}, namely $(1-X_{\mathrm{sc}}) B_{E}(T_{\mathrm{bb}}) +X_{\mathrm{sc}}I_{\mathrm{c}}(E)$ (see Sect. \ref{sec:emission_model}). 
The model was calculated with the parameters shown in Table \ref{table:params}.
} 
\label{fig:spec}
\end{figure*}

We then used the aforementioned Thomson model to obtain the final angular dependence of the radiation.
The final intensity (in the co-moving frame) was obtained jointly with \textsc{simpl} from  
\be \label{eq:tot_intensity}
I(\mu,E) = (1-X_{\mathrm{sc}})\,B_{E}(T_{\mathrm{bb}})\, f^{0}(\mu)+X_{\mathrm{sc}}\,I_{\mathrm{c}}(E)\,f(\mu,E) , 
\ee
where $f^{0}(\mu) = \mathcal{N} a^{0}(\mu)$ is the beaming for unscattered photons (normalized so that $\int_{0}^{1} \mu f^{0}(\mu) \diff \mu = 1/2$), $I_{\mathrm{c}}(E)$ is the Comptonized part of the spectrum computed with \textsc{simpl}, and 
\be \label{eq:sum_scat_orders_n-1}
f(\mu,E) = \mathcal{N}_{E}\sum_{n=1}^{N} I_{\mathrm{T}}^{n}(\mu,E),
\ee
where $\mathcal{N}_{E}$ is a normalization factor determined so that $\int_{0}^1 \mu f(\mu,E)  \diff \mu = 1/2$.
Also taking the final energy dependence of intensity directly from the Thomson model, instead of \textsc{simpl} that was used here, would not be useful because the Thomson model does not describe the observed spectra of AMPs well for most of the preferred model parameters (unlike \textsc{simpl} which we can parametrize, independently from the polarization model). 
This means that the Stokes parameters $I_{\mathrm{T}}(\mu,E)$ and $Q_{\mathrm{T}}(\mu,E)$, which were used to estimate $P(\mu,E)$,  differ from the final $I(\mu,E)$ and $Q(\mu,E) = P(\mu,E) \  I(\mu,E)$.  
However, this is not crucial, because the Thomson model is only expected to describe PD and the angular dependence of radiation with an accuracy that is good enough to obtain first order estimates for the geometrical parameters of the NS.

The modelled final angular and energy dependencies of intensity and PD (using the model parameters shown in Table~\ref{table:params}) are shown in Figs.~\ref{fig:beaming} and \ref{fig:spec}.
From the left panel in Fig.~\ref{fig:beaming}, we see that the beaming pattern shows limb darkening for lowest energies (below about 10 keV) but limb brightening down to \mbox{$\mu \approx 0.3$} for higher energies. 
The classical result of Chandrasekhar-Sobolev given by Eq.~\eqref{eq:amu_elsc},  which is shown with a black dot-dashed line, appears to be a reasonable approximation for the final angular dependence at the nominal 2--8 keV energy range of IXPE. 
On the other hand, the classical formula predicts a significantly smaller PD (in absolute value) than the optically thin Thomson model for the highest emission angles and for the most energies, as seen from the right panel of Fig.~\ref{fig:beaming}. 
We also note that assuming polarized seed photons (not shown in the figures), particularly having $p_{\mathrm{max}}=11.71\%$ in Eq.~\eqref{eq:pol_deg}, leads to even a slightly higher absolute value for the final PD of the Thomson scattered photons. 
For the highest energies (only above 18 keV), a change in the sign of the PD is also observed. 
We note that this energy depends on the parameters of the spectrum (seed photon temperature, electron temperature) and it is smaller for smaller $T_{\mathrm{bb}}$ and $T_{\mathrm{e}}$. 
For the smallest energies, PD is almost independent of energy at a given emission angle, as is also seen in the right panel of Fig.~\ref{fig:spec}. 
From the left panel of Fig.~\ref{fig:spec}, we see that the emergent intensity spectrum is close to that obtained directly from \textsc{simpl}, although the spectrum emitted at highest zenith angles has a smaller blackbody component.

\begin{figure*}
\centering
\includegraphics[width=16cm]{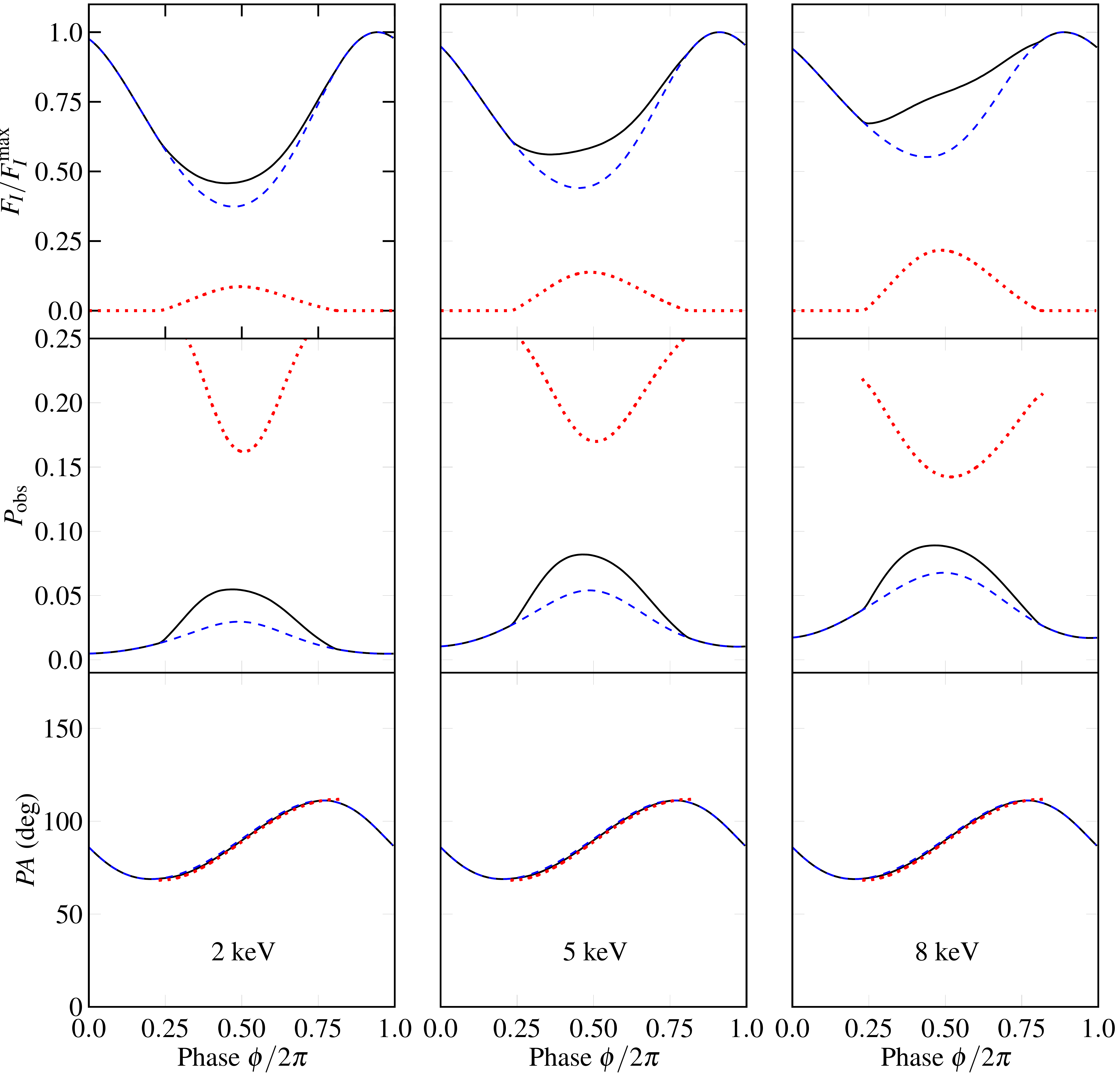}
\caption{Theoretical pulse profiles of the observed flux, PD, and PA for two antipodal spots shown for three different energies (2, 5, and 8~keV).
Solid black curves correspond to the total flux (or PD or PA), while the blue dashed curves are for the primary spot, and red dotted curves are for the secondary spot.
The parameters of the model are given in Table \ref{table:params}. 
}
\label{fig:model_new}
\end{figure*}

\begin{figure*}
\centering
\includegraphics[width=16cm]{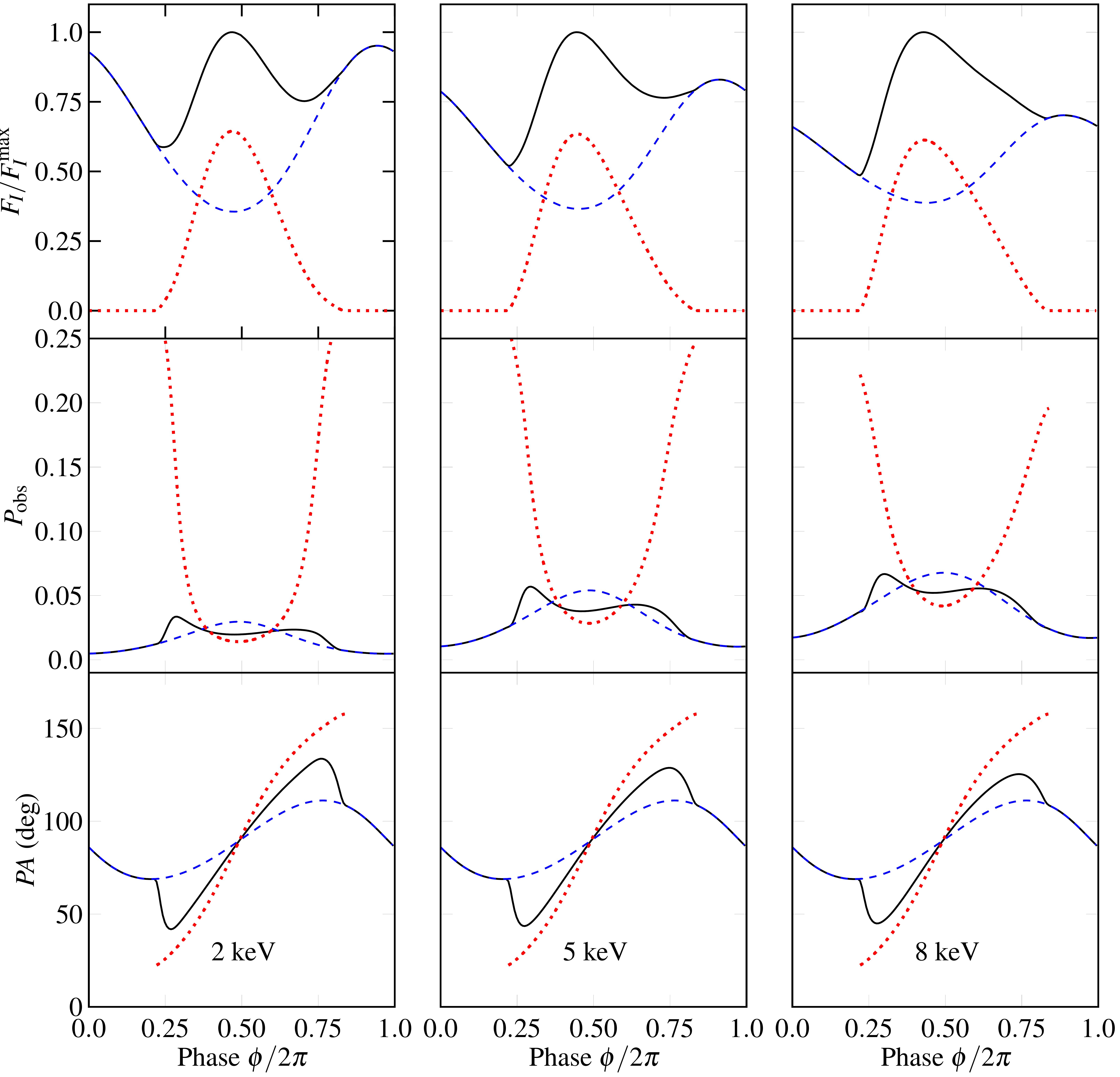}
\caption{
Same as Fig. \ref{fig:model_new}, but for a non-antipodal spot configuration with $\theta = 20 \degr$,  $\theta_{2} = 120 \degr$, and $\phi_{12} = 180 \degr$. }
\label{fig:model_non-antipod}
\end{figure*}

\subsubsection{Pulse profiles}\label{sec:pulse_model}

When calculating pulse profiles, PD $P(\mu,E)$ and the beaming of the Comptonized spectral component $f(\mu,E)$ were obtained by interpolating pre-computed tables. 
Following the methods of \citet{VP04} and \citet{LSNP20} to transform the quantities from the spot frame to the observer frame, we obtained the Stokes flux vector ($F_{I}$, $F_{Q}^{\mathrm{pul}}$, $F_{U}^{\mathrm{pul}}$) as a function of observed phase $\phi$.
The observed phase differs from the pulsar rotation phase $\phi_{\mathrm{rot}}$ due to the different time delays for different sub-spots at different phases. 
However, unlike in \citet{VP04} and \citet{LSNP20}, we used the equatorial radius as the reference radius and zero as the reference impact parameter when computing the delays (as in \citealt{PB06} and \citealt{SNP18}).  
The rotational phase of the pulsar was set to zero when the primary spot was closest to the observer.

The obtained Stokes parameters are defined in the polarization basis related to the projection of the pulsar rotation axis on the sky. 
Therefore, we also decided to take the position angle $\chi$ of the pulsar rotation axis into account  by correcting the computed Stokes parameters using the following equations:  
\beq \label{eq:chi_def_}
F^{\mathrm{mod}}_{Q} = F_{Q}^{\mathrm{pul}}\cos(2 \chi)-F_{U}^{\mathrm{pul}}\sin(2\chi), \nonumber \\
F^{\mathrm{mod}}_{U} = F_{Q}^{\mathrm{pul}}\sin(2 \chi)+F_{U}^{\mathrm{pul}}\cos(2\chi),
\eeq
where $F^{\mathrm{mod}}_{Q}$ and $F^{\mathrm{mod}}_{U}$ are the final phase-resolved model Stokes spectra.
In our simulated data, we assumed that $\chi = 0$, but we kept it  as a free parameter when calculating parameter constraints. 
Finally, the observed PD was obtained as
\beq \label{eq:Pobs}
P_{\mathrm{obs}} = \frac{\sqrt{(F_{Q}^{\mathrm{mod}})^2+(F_{U}^{\mathrm{mod}})^2}}{F_{I}}. 
\eeq

We now consider the case with one or two antipodal spots.  
The modelled polarized pulses, in the case of model 8 presented in Table~\ref{table:all_models}, are shown in Fig.~\ref{fig:model_new} for three different observed photon energies (2, 5, and 8 keV). 
The contributions from primary and secondary spots are indicated with blue dashed and red dotted curves correspondingly (the blue curve represents the fiducial model shown in Table~\ref{table:params}). 
We see only weak dependence of the observed PD $P_{\mathrm{obs}}$ and PA on the energy. 
During one rotational period, all pulse profiles show only one maximum because the second spot is significantly less visible compared to the primary one.
For all the shown energies, the PD of each spot (shown in the middle row in Fig.~\ref{fig:model_new}) is highest when the spot is most inclined to the observer, but it is still visible. 
Also, the maximum PD is significantly higher if considering emission from two spots. 
At energies above 10 keV, the sign of $P$ changes (see Fig.~\ref{fig:spec}). 
This results in a jump in PA by 90\degr\ at certain phases. 
The pulsar phase bin where such a jump occurs depends on the energy. 
However, we only used the model in the range of 2--8 keV, where the current instrumental response of the IXPE detector is best validated.

\begin{figure*}
\centering
\includegraphics[width=8cm]{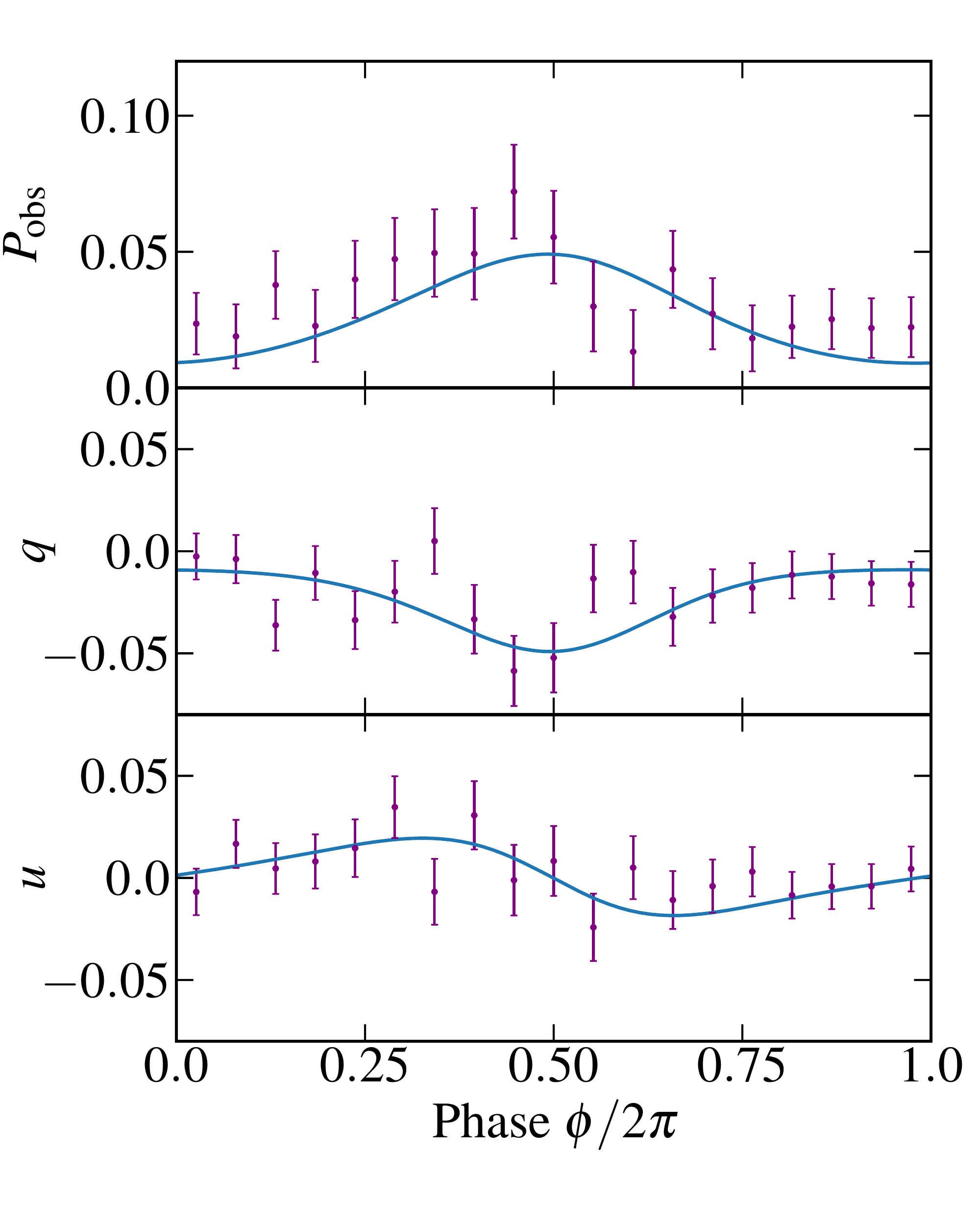}
\hspace{0.5cm}
\includegraphics[width=8cm]{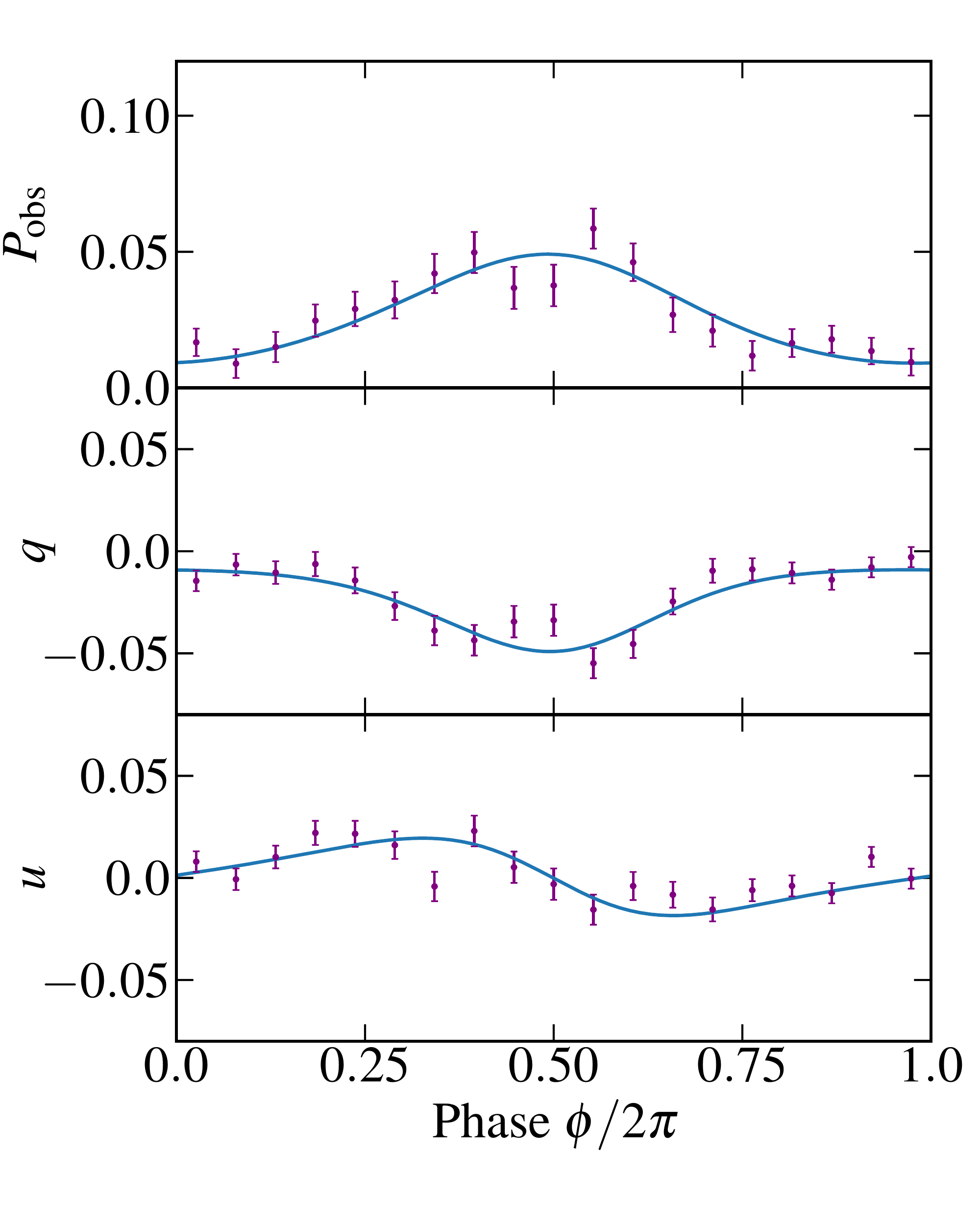}
\caption{Simulated PD and normalized Stokes $q$ and $u$ profiles for 200 ks (\textit{left panels}) and 1 Ms (\textit{right panels}) exposure times and the fiducial one-spot model when computing a weighted average over energy range 2--8 keV. 
The blue curve shows the theoretical model and the purple dots show the simulated observed data including the measurement errors. 
}
\label{fig:data_qu200ks_1000ks}
\end{figure*}

We now present our model in the case of a non-antipodal spot configuration. 
We assumed that the co-latitudes of the spots are $\theta = 20 \degr$ and $\theta_{2} = 120 \degr$ with the difference in longitude between the spots being $\phi_{12} = 180 \degr$ (model 9 presented in Table~\ref{table:all_models}). 
We show the model in Fig. \ref{fig:model_non-antipod}. 
The contribution from the primary spot is the same as in Fig. \ref{fig:model_new}, but now the secondary spot is more visible to the observer. 
The observed PD is lower because photons reaching the observer are emitted to smaller angles relative to the spot normal. 
We also detect abrupt breaks in PA at those phases when the secondary spot appears and disappears because the non-antipodal spot has a largely different PA (as also discussed by \citealt{LSNP20}).

\subsection{IXPE data simulation}\label{sec:ixpeobssim}

We used the IXPE observation-simulation framework \textsc{ixpeobssim} version 12.0.0~\citep{ixpeobssim2019} to generate our synthetic data. 
The programme is designed to fold a complete source model (described in Sect. \ref{sec:pol_modelling}, in our case) with the current best estimate of the instrument response functions to produce simulated event files (also known as event or photon lists) for a given observation time.
The event files are essentially identical in format to those that will be produced from real observations and they include all the relevant physical properties of the events, such as the measured arrival time, energy, sky-position, and, most notably, the photoelectron azimuthal angle $\psi_{k}$. 
The latter is key to the polarization measurement and is the basic ingredient for the analysis described in this section.

Starting from a given photon list, we built the so-called Stokes parameter spectra $C_{Q}^{\mathrm{obs}}$ and $C_{U}^{\mathrm{obs}}$. 
These are essentially weighted histograms of detector counts, binned in energy, where the weights are dictated by the reconstructed azimuthal angle of each event, $q_{k} = \cos 2 \psi_{k}$ and $u_{k} = \sin 2 \psi_{k}$, according to the formalism described in \citet{KCB15}. 
The statistical errors were propagated by summing the weights themselves in quadrature, according to the standard formalism of weighted histograms.
From a practical standpoint, these data products were stored in FITS files whose format conforms to the standard OGIP PHA type I file format in order to be interoperable with the high-level analysis tools used by the X-ray community, such as {\sc xspec}.
The Stokes spectra were created separately for each of the 19 pulsar phase bins.

All the IXPE instrument response functions and associated binned products are defined between 1 and 12 keV in 275 equal-width energy bands. 
However, we only used channels in the nominal 2--8~keV IXPE energy band to avoid any complication from the complexity of the detector response outside this range. 
The effective area below 2 keV and above 8 keV is so small that this has effectively very little practical implications for the analysis.

The observed Stokes parameters (and their measured errors) were transformed to polarization estimates (directly comparable with the input model) by dividing them by the effective modulation factor of each observed energy channel, $C_{Q} = C_{Q}^{\mathrm{obs}}/\mu_{\mathrm{eff}}$ and $C_{U} = C_{U}^{\mathrm{obs}}/\mu_{\mathrm{eff}}$ \citep[as explained in][]{KCB15}.
In addition, we simulated the data in terms of the normalized Stokes profiles $q=C_{Q}/C_{I}$ and $u=C_{U}/C_{I}$, where $C_{I}$ is the observed count spectrum (see simulated data presented in Sect.~\ref{sec:sim_data}). 
These can be compared to the normalized Stokes parameters $q_{\mathrm{m}}=F^{\mathrm{mod}}_{Q}/F_{I}$ and $u_{\mathrm{m}}=F^{\mathrm{mod}}_{U}/F_{I}$, which can be predicted directly from the theoretical model, because the ratio of two Stokes parameters is not expected to be sensitive to the energy response of the detector.

When estimating the measurement errors and comparing the data to the model (in Sect. \ref{sec:sim_data}), we also combined observations from all the energy channels between 2 and 8 keV. 
The broadband values were computed as averages of $q$ and $u$, weighted with the energy-dependent effective area according to the formalism in \citet{KCB15}, and they were compared to the modelled broadband values of $q_{\mathrm{m}}$ and $u_{\mathrm{m}}$.
When analysing the data using the Bayesian inference (in Sect. \ref{sec:param_con}), we used the same technique in order to speed up the computation, and we created eight logarithmic energy bands between 2 and 8 keV, which were independently fitted.

\subsection{Bayesian modelling}\label{sec:bayesian}

The obtained Stokes profiles were fitted using an affine invariant ensemble sampler with the \textsc{emcee} package of \textsc{python} \citep{emcee}.
The posterior probability densities were calculated to the free parameters of our model.
For most simulations, these were the observer inclination $i$, spot co-latitude $\theta$, phase shift $\Delta \phi$, and the position angle $\chi$ of the pulsar spin axis. 
In the case of the model with a large spot size ($\rho=30\degr$), we also kept the size of the spot $\rho$ as a free parameter; furthermore, in the case of the two-spot models, we also kept the co-latitude $\theta_{2}$ of the secondary spot and the longitude difference $\phi_{12}$ between the spots as free parameters.
The correct values of these and also the fixed model parameters are shown in Table~\ref{table:params} (for the fiducial model) and in Table~\ref{table:all_models} (for other models). 

The synthetic normalized Stokes $q$ and $u$ data (created using \textsc{ixpeobssim} as described in Sect. \ref{sec:ixpeobssim}) were fitted against the modelled $q_{\mathrm{m}}$ and $u_{\mathrm{m}}$ assuming that the probability densities of $q$ and $u$ are uncorrelated and normally distributed around $q_{\mathrm{m}}$ and $u_{\mathrm{m}}$ with measured errors as the standard deviation. 
As mentioned in Sect. \ref{sec:ixpeobssim}, by fitting normalized Stokes values, we improved the computational efficiency as we did not need to transform every model to observed data using a forward-folding approach (including the response of the detector).
The likelihoods for a given model were calculated separately for 19 phases and eight energy bins of the simulated data (where the energy channels were combined as explained in Sect. \ref{sec:ixpeobssim}) and summed to get the total likelihood. 
The prior probability distributions of all the sampled parameters were assumed to be uniform and the limits of the priors were set to (0\degr, 90\degr) in $i$ and $\theta$, ($-$90\degr , 90\degr) in $\chi$, ($-$180\degr , 180\degr) in $\Delta \phi$, (0\degr , 60\degr) in $\rho$, (0\degr , 180\degr) in $\theta_{2}$, and (0\degr , 360\degr) in $\phi_{12}$.

\begin{figure}
\centering
\includegraphics[width=8.0cm]{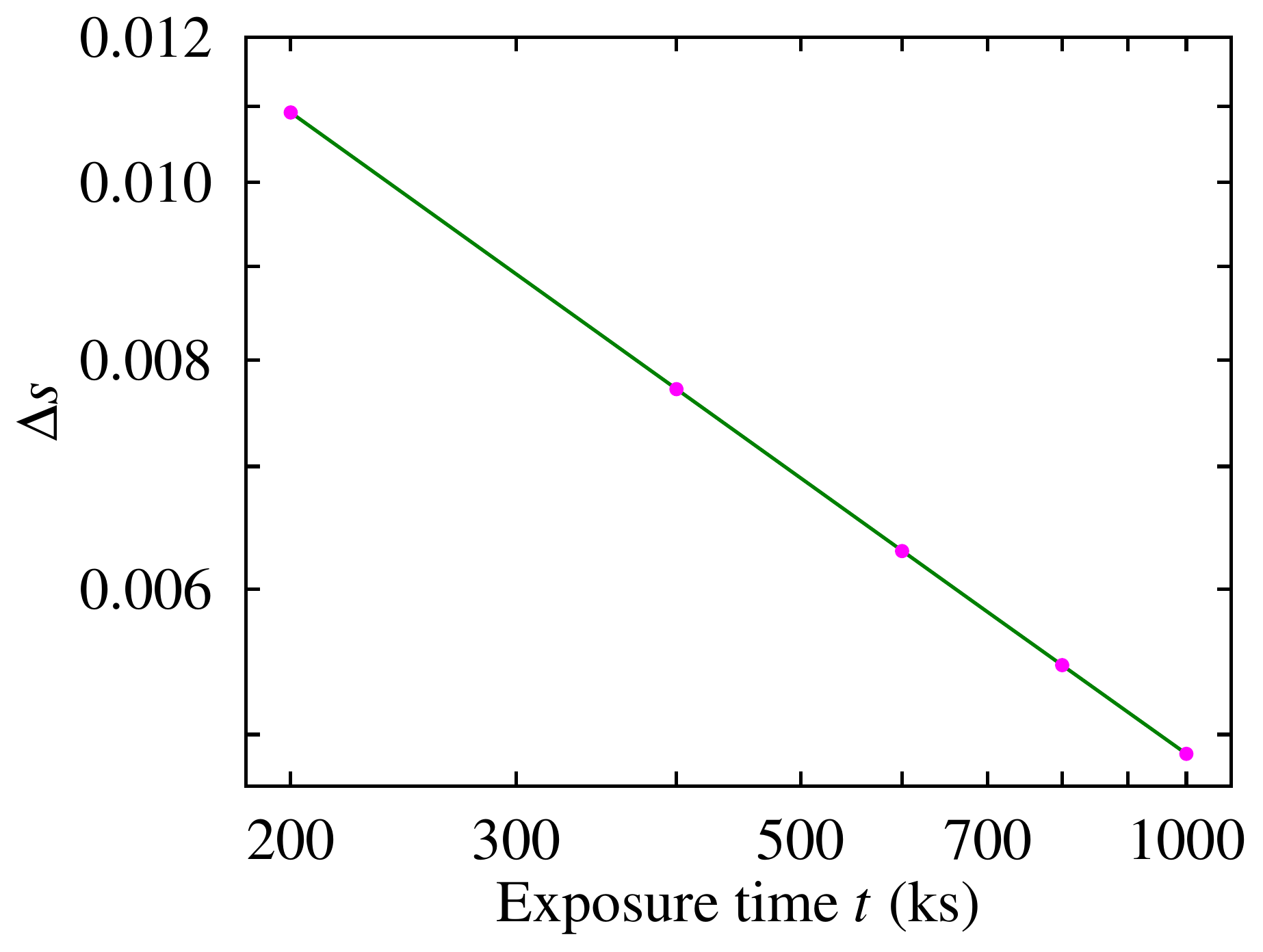}
\caption{
Average absolute minimum errors in the measured energy-integrated Stokes $q$ and $u$ as a function of observation time (for the one-spot  models specified in Table \ref{table:all_models}). 
The curves were calculated for 200, 400, 600, 800, and 1000
ks (shown as purple dots) and interpolated linearly in a logarithmic scale for other values. 
}
\label{fig:data_errors_abs}
\end{figure}

\begin{figure}
\centering
\includegraphics[width=8.0cm]{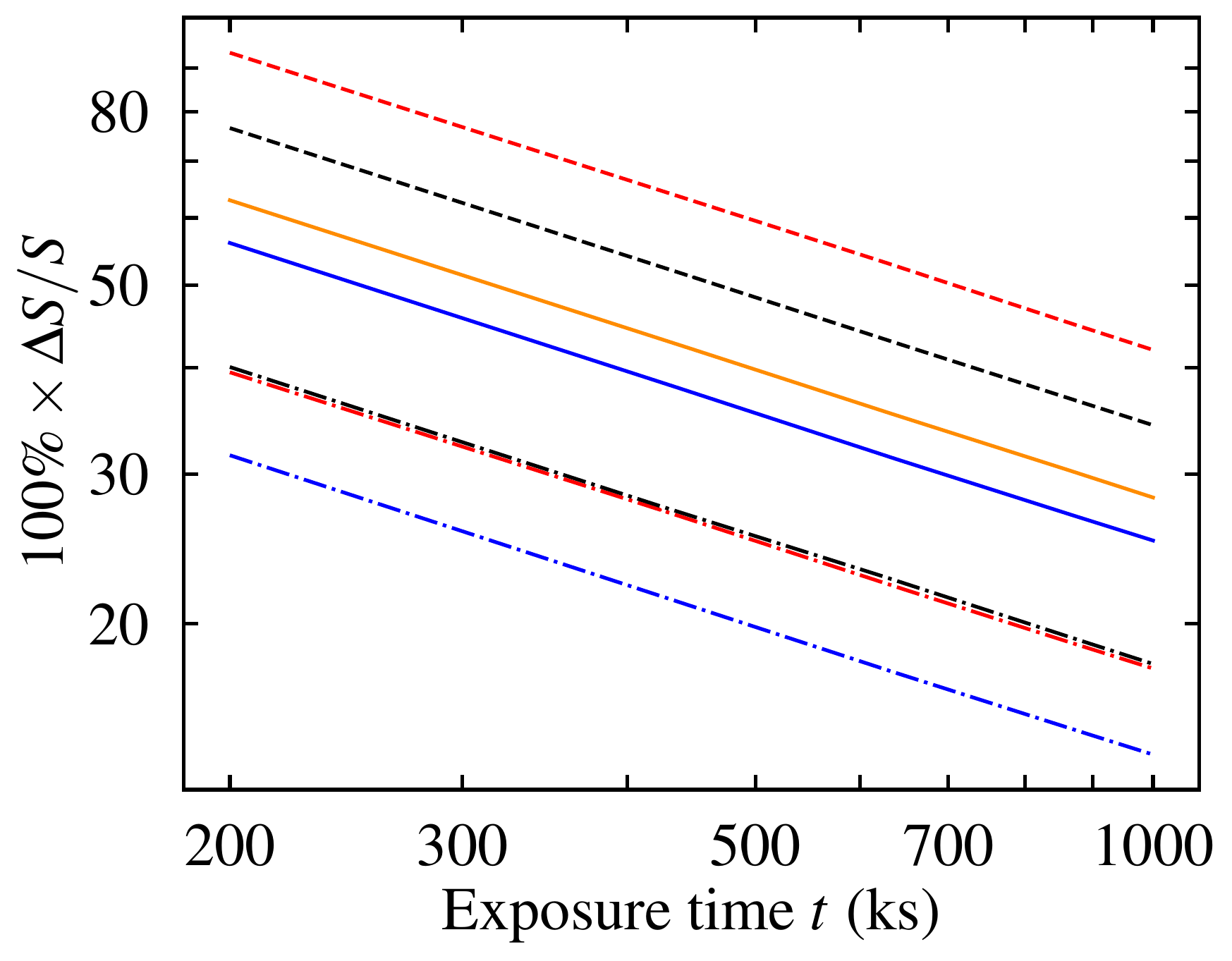}
\caption{
Average relative minimum errors in the measured energy-integrated Stokes $q$ and $u$ as a function of observation time for the one-spot models (calculated as $\Delta S /S = ([\Delta q / q_{\mathrm{m}}]_{\mathrm{min}} + [\Delta u / u_{\mathrm{m}}]_{\mathrm{min}})/2$. 
The solid blue curve shows the errors with the fiducial parameters (shown in Table \ref{table:params}).
The blue dash-dotted curve corresponds to similar data but with $p_{\mathrm{max}}=0.1171$,  the black dashed curve with $i=50 \degree$, the black dash-dotted curve with $i=70\degree$, the red dashed curve with $\theta=10 \degree$, the red dash-dotted curve with $\theta = 30 \degree$, and the orange solid curve with $\rho = 30 \degree$ (also explained in Table \ref{table:all_models}). 
Curves were interpolated from the calculated points as in Fig. \ref{fig:data_errors_abs}. 
}
\label{fig:data_errors_rel}
\end{figure}

\begin{figure*}
\centering
\includegraphics[width=8cm]{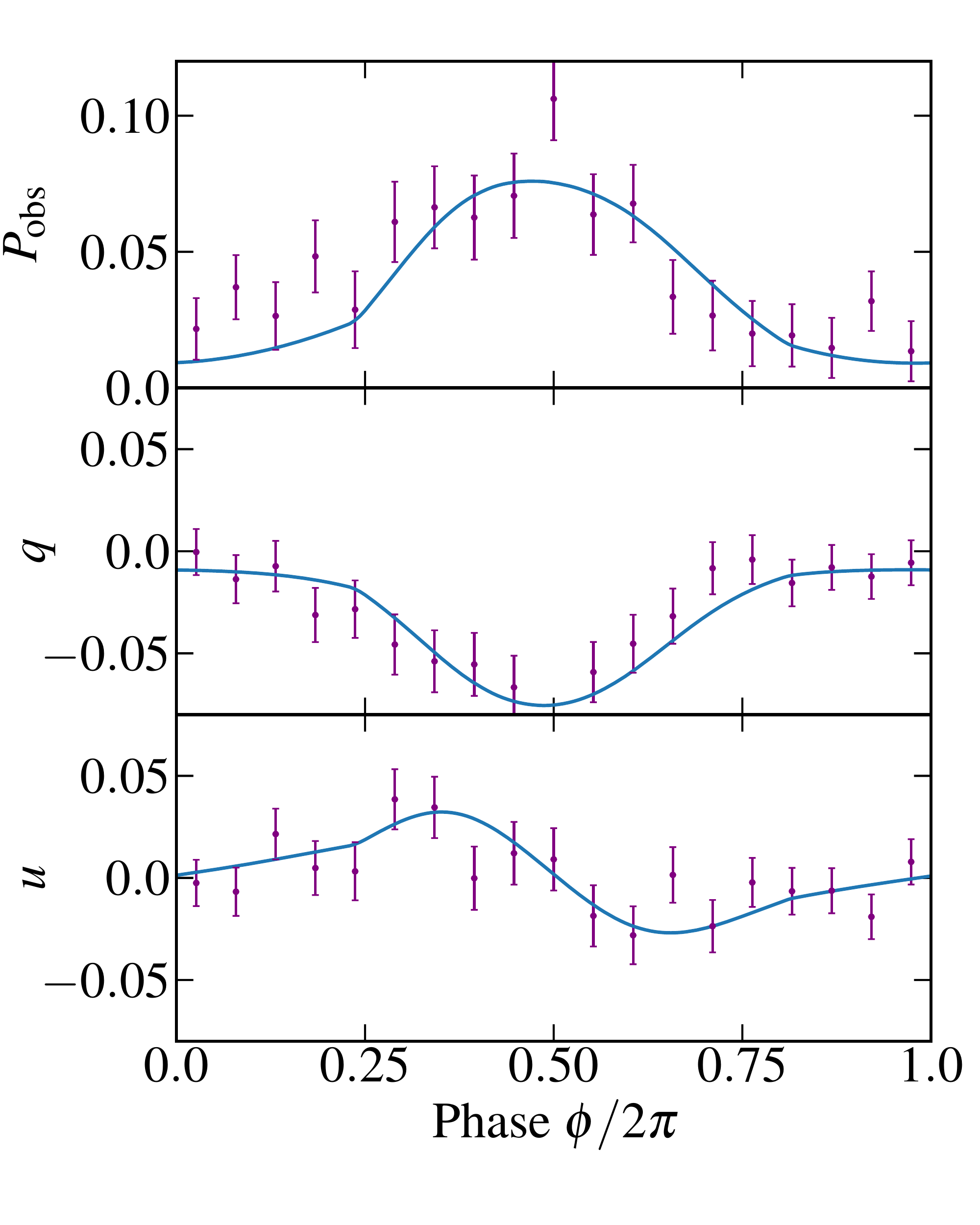}
\hspace{0.5cm}
\includegraphics[width=8cm]{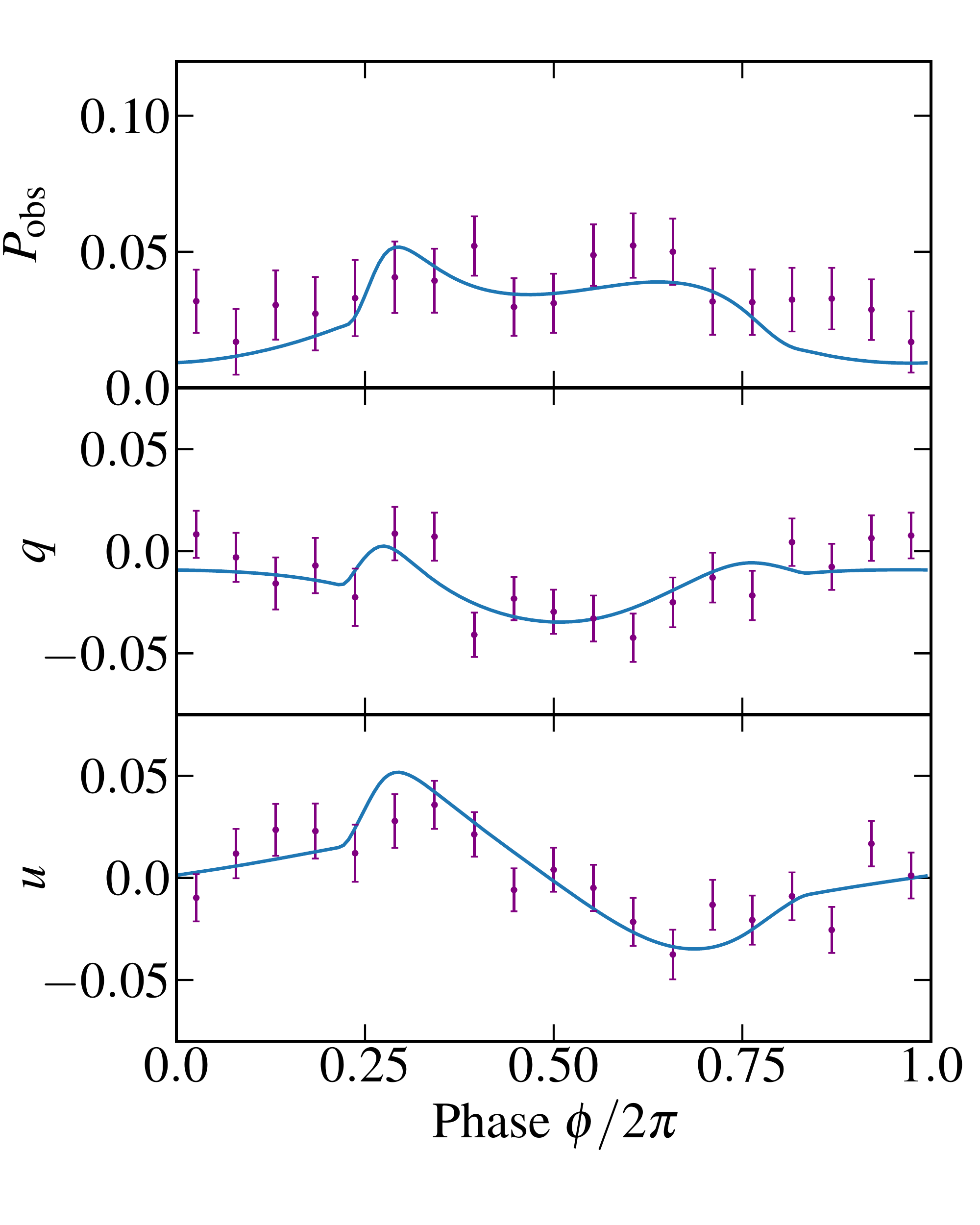}
\caption{
Simulated polarization data as in Fig. \ref{fig:data_qu200ks_1000ks} (for 200 ks exposure time), but for the case of two antipodal spots with  $\theta = 20 \degr$ and $\theta_{2} = 160 \degr$ (\textit{left panels}), and two non-antipodal spots with $\theta = 20 \degr$ and $\theta_{2} = 120 \degr$(\textit{right panels}). 
}
\label{fig:data_2spots}
\end{figure*}

\section{Results}\label{sec:results}

\subsection{Simulated data}\label{sec:sim_data}

We now\ present the simulated data and estimated measurement accuracy obtained for the one-spot models with different exposure times and slightly different model parameters. 
The computations were done with exposures of 200, 400, 600, 800, and 1000~ks.
In all of our simulations, we assumed a 100~mCrab source, corresponding to approximately \source\ during its peak luminosity.
For other sources, which are less bright, a longer observation time is expected to be required to obtain similar results.
On the other hand, a higher PD than our assumption (about a maximum 5\% observed PD for our fiducial parameter set) should affect the simulated data in the same way as increasing the observation time.

The simulated broadband PD and Stokes $q$ and $u$ profiles for the shortest (200 ks) and longest (1000 ks) exposure times, in the case of one observation realisation of the fiducial parameter set, are shown in Fig.~\ref{fig:data_qu200ks_1000ks}. 
We see that the produced data are not biased from the theoretical model shown in blue. 
The measurement errors are also significantly smaller in the case of a longer exposure time. 

The errors as a function of the observation time are examined in more detail, for all calculated one-spot models and exposure lengths, in Figs.~\ref{fig:data_errors_abs} and \ref{fig:data_errors_rel}, where we present the average minimum errors (absolute and relative, respectively) in $q$ and $u$ as a function of observation time $t$.
From these figures, we see that increasing the observation time from 200 to 1000 ks improves the accuracy in the measured $q$ and $u$ from 30--90\%  to 15--40\% depending strongly on the model parameters. 
Of course, the exact values for the errors also depend on the adopted number of phase bins and they are only relevant when comparing and analysing data with a similar setup. 

In any case, the presented errors scale as $1/\sqrt{t}$, which is as expected. 
The relative errors in Fig. \ref{fig:data_errors_rel} are smallest for those models that produce the highest PD. 
These are the models where the initial polarization is non-zero (blue dash-dotted curve), and where the inclination or co-latitude is higher than in the fiducial model (black and red dash-dotted curves, respectively). 
Larger errors are expected for the case with a larger spot size (solid orange curve) and smaller inclination or co-latitude (black and red dashed curves). 
On the contrary, absolute errors in Fig. \ref{fig:data_errors_abs} are almost identical in all models because they are only determined by the total amount of observed counts and the brightness of the source is fixed in every model. 

We also simulated polarization data detected assuming either two antipodal or non-antipodal spots (corresponding to the two-spot models shown in Figs. \ref{fig:model_new} and \ref{fig:model_non-antipod}), but using only a single observation realisation with a 200 ks observation time.  
The results are shown in Fig. \ref{fig:data_2spots}, where we see that the measured broadband PD and Stokes parameters for the antipodal case significantly differ from those of the non-antipodal case. 
Relative errors in the measured Stokes parameters are slightly higher for the non-antipodal case because the PD is smaller for the chosen model parameters.

\begin{figure*}
\centering
\includegraphics[width=8.2cm]{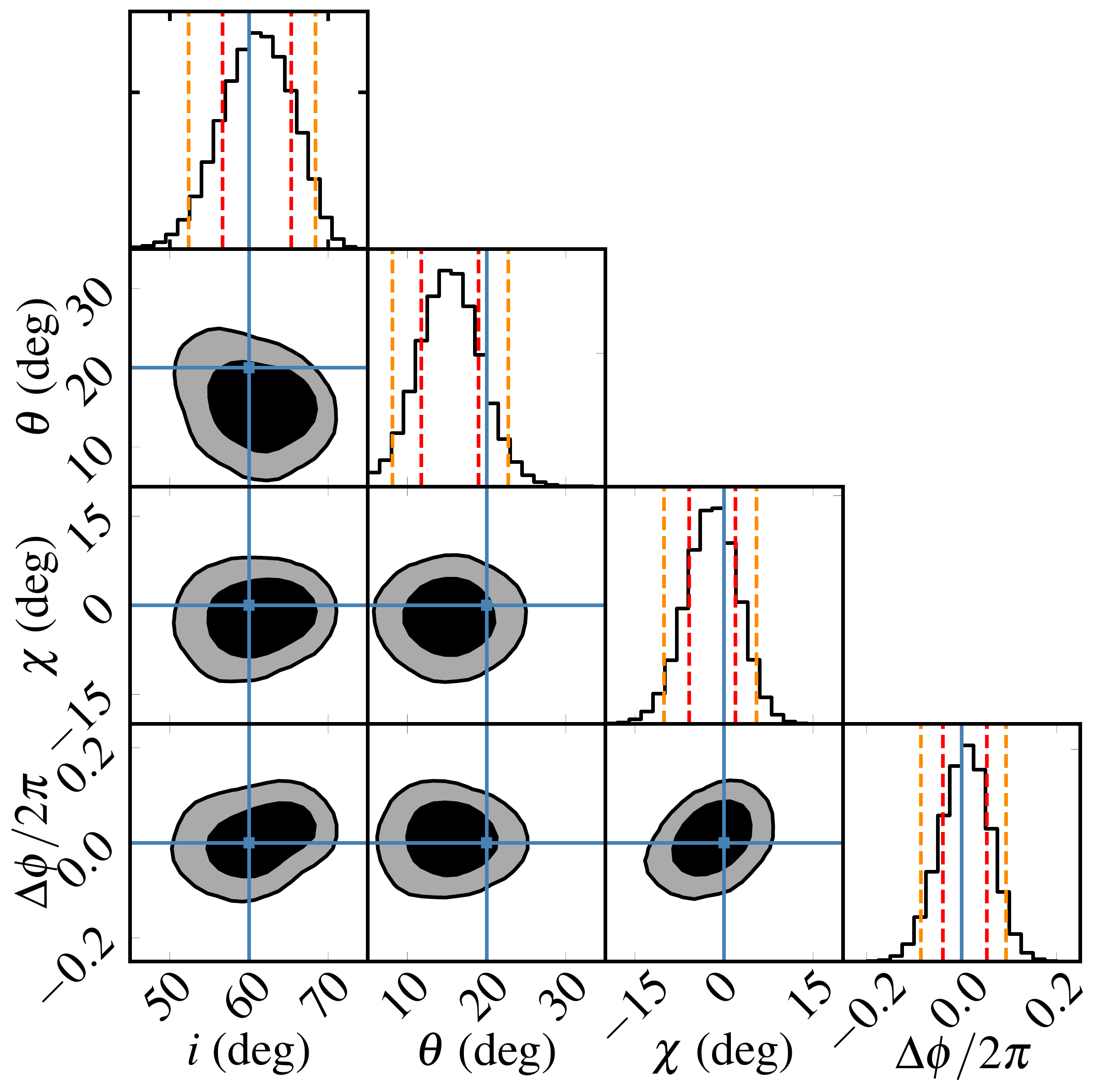}
\hspace{0.5cm}
\includegraphics[width=8.2cm]{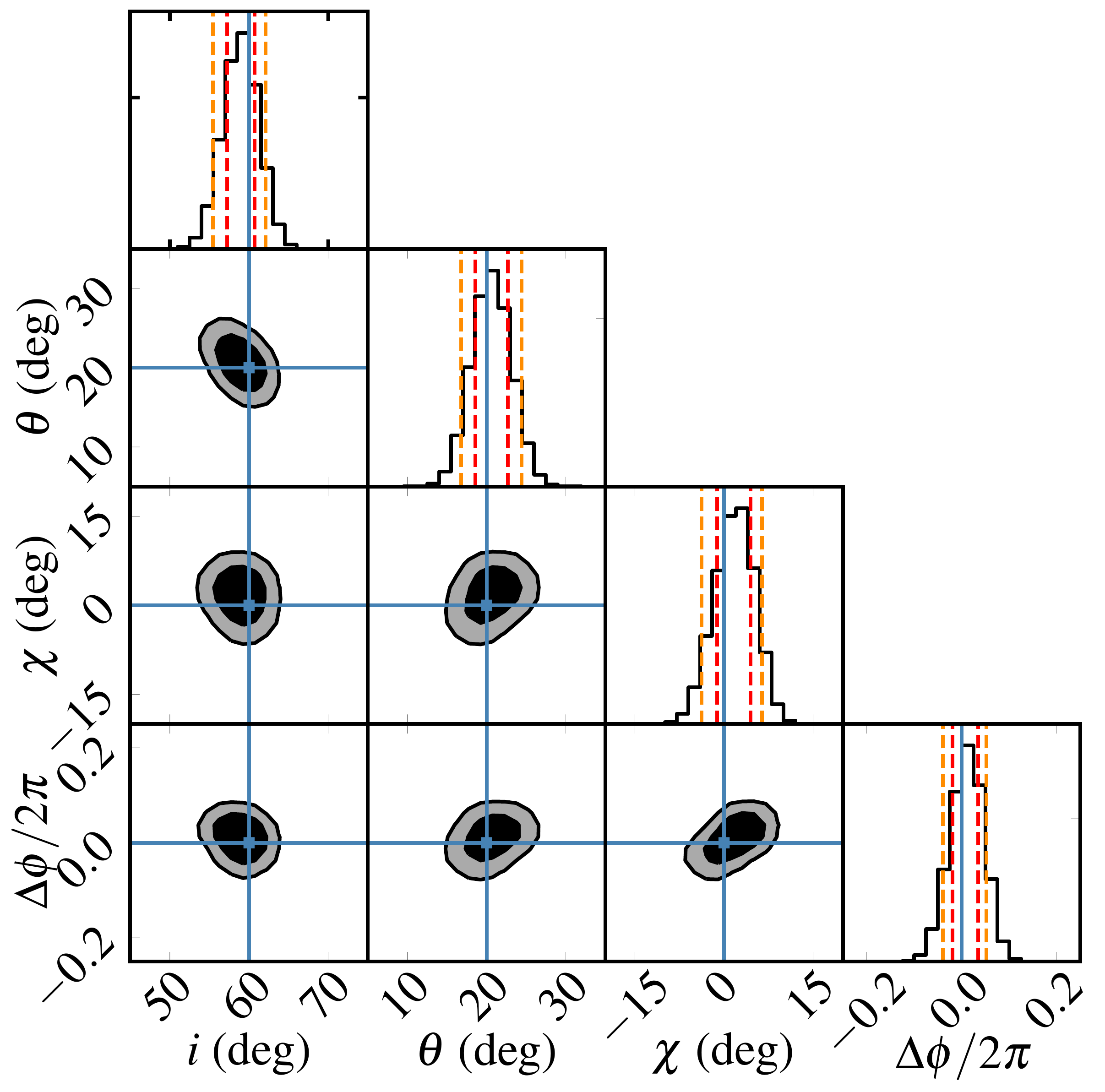}
\caption{
Posterior probability distributions for NS parameters when combining fits from three realisations of synthetic data produced assuming a 200 ks (\textit{left panels}) and a 1 Ms (\textit{right panels}) exposure time  and the fiducial model shown in Table \ref{table:params}. 
In the two-dimensional posterior distributions, the black colour shows a 68\% and the grey colour shows a 95\% highest posterior density credible region. 
In the one-dimensional posterior distributions, the red dashed lines show a 68\% and the dark orange dashed lines show a 95\% highest posterior density credible interval. 
The blue lines show the input values. 
}
\label{fig:mcmc_200ks_1000ks}
\end{figure*}

\begin{table*}
  \caption{Most probable values and $68\%$ and $95\%$ credible intervals for five different simulations averaged from three realisations of the synthetic data produced using the fiducial model.}
\label{table:conflimits}
\centering
  \begin{tabular}[c]{l c c c c c}
    \hline\hline
      Quantity & $95\%$ lower limit & $68\%$ lower limit & Most probable value& $68\%$ upper limit & $95\%$ upper limit \\ \hline   
\multicolumn{6}{c}{Exposure 200\,ks} \\
$i$ ($\deg$) & $54.0$ & $57.6$ & $60.8$ & $63.9$ & $66.3$ \\
$\theta$ ($\deg$) & $9.09$ & $12.2$ & $15.7$ & $19.3$ & $23.2$ \\
$\chi$ ($\deg$)& $-9.6$ & $-5.9$ & $-1.9$ & $2.0$ & $5.8$ \\
$\Delta \phi/2\pi$ & $-0.07$ & $-0.03$ & $0.007$ & $0.04$ & $0.09$ \\ \hline
\multicolumn{6}{c}{Exposure 400\,ks} \\
$i$ ($\deg$) & $52.1$ & $55.2$ & $57.6$ & $60.0$ & $62.0$ \\
$\theta$ ($\deg$) & $13.6$ & $16.0$ & $18.7$ & $21.2$ & $23.8$ \\
$\chi$ ($\deg$)& $-3.3$ & $-0.40$ & $2.7$ & $5.9$ & $9.0$ \\
$\Delta \phi/2\pi$ & $-0.03$ & $0.002$ & $0.03$ & $0.05$ & $0.08$ \\ \hline
\multicolumn{6}{c}{Exposure 600\,ks} \\
$i$ ($\deg$) & $53.8$ & $56.2$ & $58.3$ & $60.1$ & $61.9$ \\
$\theta$ ($\deg$) & $15.9$ & $18.0$ & $20.1$ & $22.2$ & $24.5$ \\
$\chi$ ($\deg$)& $-5.1$ & $-2.7$ & $-0.42$ & $2.1$ & $4.9$ \\
$\Delta \phi/2\pi$ & $-0.04$ & $-0.02$ & $-0.003$ & $0.02$ & $0.03$ \\ \hline
\multicolumn{6}{c}{Exposure 800\,ks} \\
$i$ ($\deg$) & $56.5$ & $58.5$ & $60.3$ & $62.0$ & $63.3$ \\
$\theta$ ($\deg$) & $15.7$ & $17.4$ & $19.1$ & $21.0$ & $22.9$ \\
$\chi$ ($\deg$)& $-3.4$ & $-1.4$ & $0.43$ & $2.3$ & $4.2$ \\
$\Delta \phi/2\pi$ & $-0.04$ & $-0.03$ & $-0.01$ & $0.004$ & $0.02$ \\ \hline
\multicolumn{6}{c}{Exposure 1000\,ks} \\
$i$ ($\deg$) & $56.0$ & $57.6$ & $59.3$ & $60.7$ & $62.0$ \\
$\theta$ ($\deg$) & $17.2$ & $18.8$ & $20.4$ & $22.2$ & $23.8$ \\
$\chi$ ($\deg$)& $-2.0$ & $-0.22$ & $1.5$ & $3.5$ & $5.2$ \\
$\Delta \phi/2\pi$ & $-0.02$ & $-0.003$ & $0.01$ & $0.02$ & $0.04$ \\ \hline

  \end{tabular}
\tablefoot{The quantities shown in the table are the observer inclination $i$, spot co-latitude $\theta$, position angle $\chi$ of the pulsar spin axis, and the phase shift $\Delta \phi$. 
The correct values of these parameters are $i = 60 \degree$, $\theta = 20 \degree$, $\chi = 0 \degree$, and $\Delta \phi = 0 \degree$. 
}  
\end{table*}

\subsection{Parameter constraints}\label{sec:param_con}

We now present the parameter constraints obtained when fitting synthetic data and start again with the one-spot models.
The resulting posterior probability distributions for the model parameters are shown in Fig.~\ref{fig:mcmc_200ks_1000ks} for the fiducial data with the shortest and longest observation times (200 and 1000 ks). 
Because a slight variation in the constraints was detected for different realisations of the same synthetic data, we show the combined posterior distributions for three different realisations of the data (including those shown in Fig.~\ref{fig:data_qu200ks_1000ks}). 
From Fig.~\ref{fig:mcmc_200ks_1000ks}, we see that a minor bias in spot co-latitude $\theta$ appears with a 200 ks observation, but no biases arise in the case of a 1000 ks observation.
All the free parameters, which are the observer inclination $i$, spot co-latitude $\theta$, position angle $\chi$, and the phase shift $\Delta \phi,$ are clearly constrained better than their prior limits (given in Sect.~\ref{sec:bayesian}), and they show no strong correlation with each other. 

The results also appear similar in the case of the other observation lengths and calculated models, although with a varying magnitude of the credible intervals. 
One-dimensional posterior results averaged over the three different synthetic data generations of the fiducial model are also shown in Table~\ref{table:conflimits}. 
Furthermore, in the case of the simulations where the spot size was a free parameter (those where the correct value was $\rho=30\degr$), no stringent constraints were found to the spot size.

\begin{figure}
\centering
\includegraphics[width=8cm]{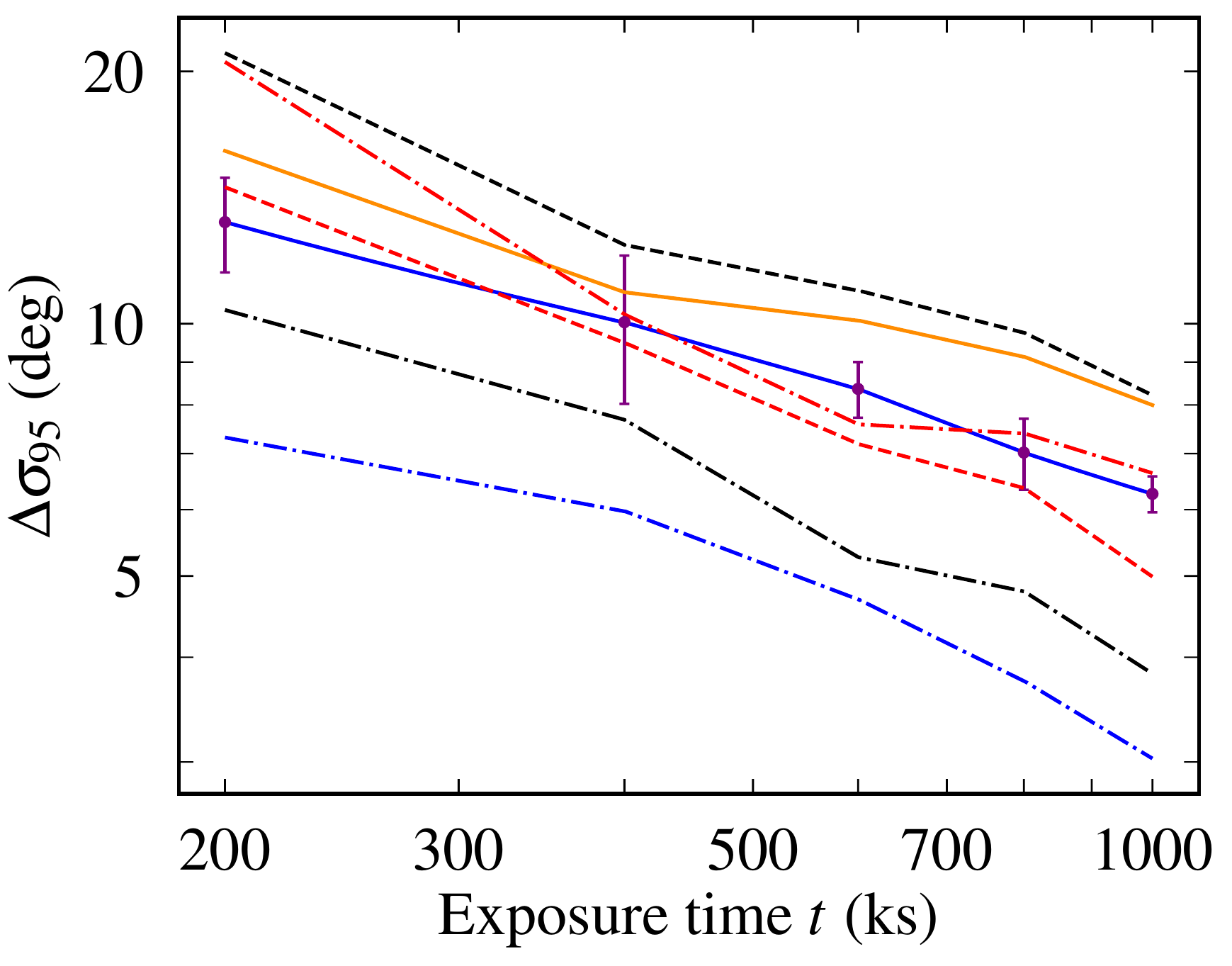}
\caption{
Average width of the $95\%$ credible interval for the measured $i$ and $\theta$ as a function of the observing time.
The curves, colours, and interpolation method correspond to those in Fig. \ref{fig:data_errors_rel}, 
but the solid blue curve corresponds to the average constraints from the three different synthetic data generations (using the same model parameters). 
Purple error bars show the maximum deviation between the average and single data generation results. 
The other lines show constraints for only one realisation of the synthetic data (for each exposure time).
}
\label{fig:data_errors_sigma}
\end{figure}

In Fig.~\ref{fig:data_errors_sigma} we illustrate how the obtained average credible interval for $i$ and $\theta$ depends on the observation time for all the one-spot models, however, when using only one realisation of the data for the other models than the fiducial one. 
For the fiducial model, we see that increasing the observation time from 200 to 1000 ks improves the width of the 95\% probability interval from about $13\degr$ to $6\degr$.
We detect a slight variation between the constraints from different models and observation lengths, which can be explained by the statistical uncertainties in the generated data.\ However, ordinarily, the constraints become tighter when increasing the exposure time. 
We note that the presented constraints approximately scale as the relative errors shown in Fig.~\ref{fig:data_errors_rel}.

\begin{figure*}
\centering
\includegraphics[width=8.2cm]{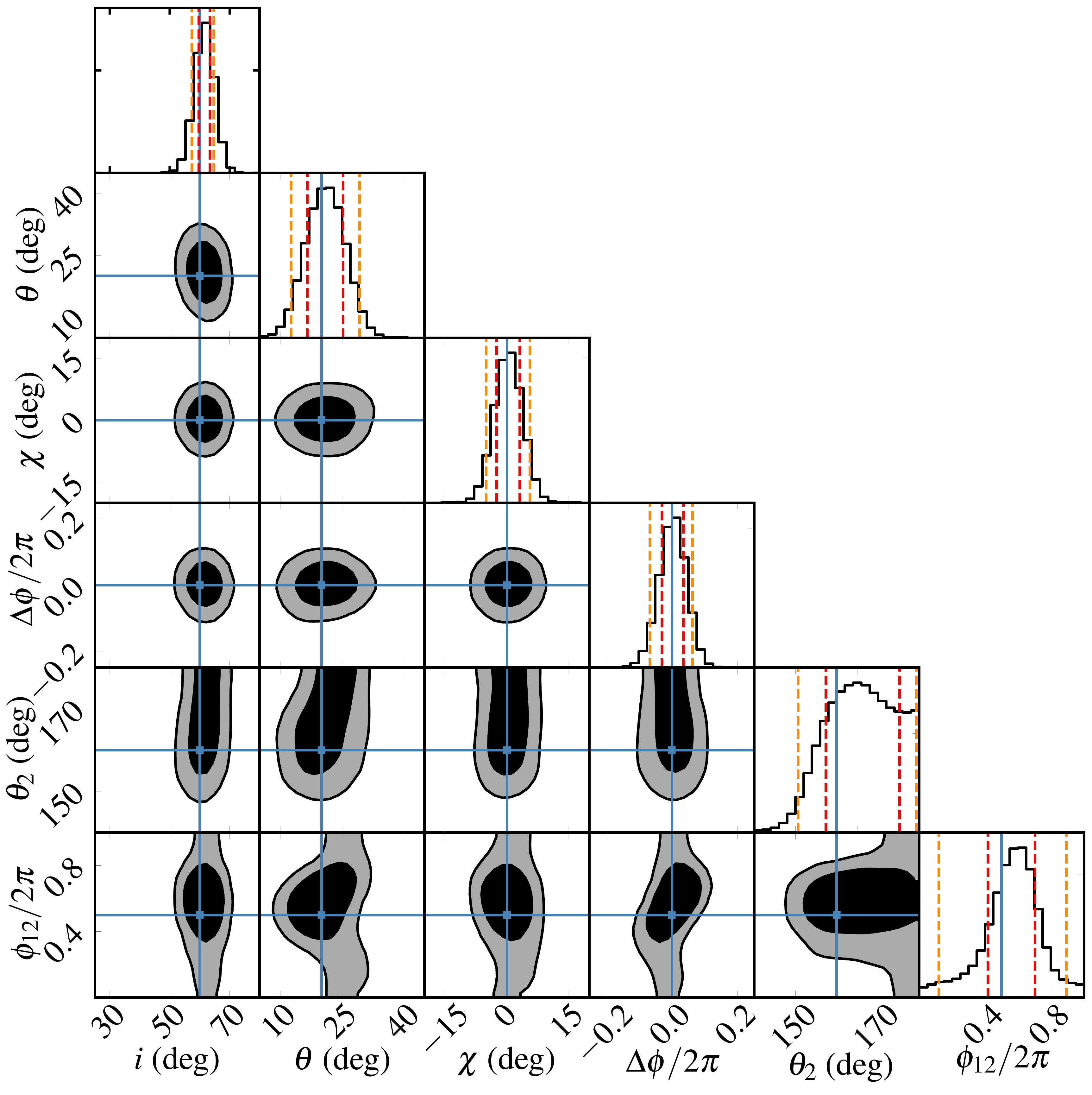} 
\hspace{0.5cm}
\includegraphics[width=8.2cm]{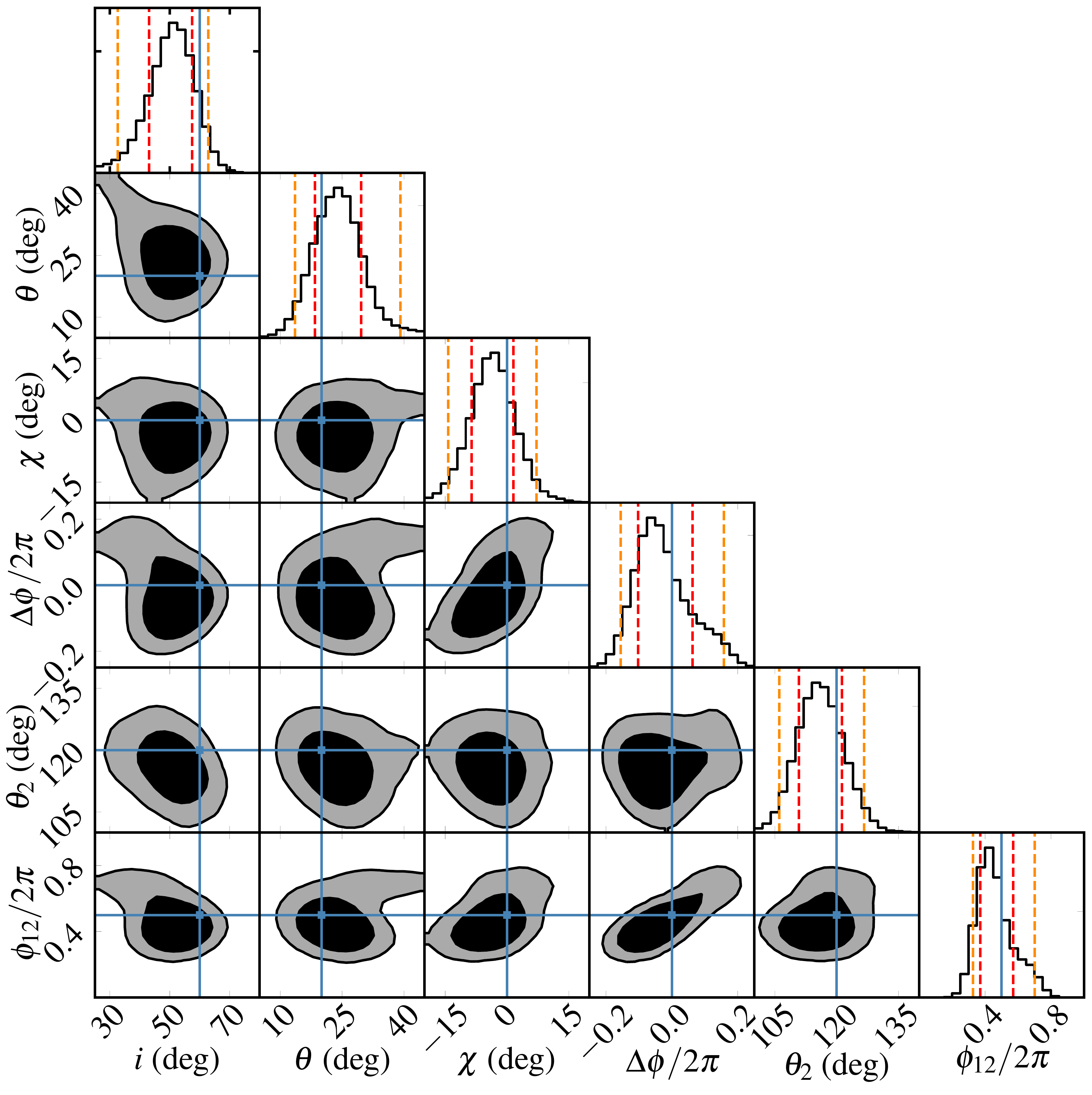}
\caption{
Posterior probability distributions for NS parameters similar to the left panel of Fig.~\ref{fig:mcmc_200ks_1000ks}, but for one data generation with two antipodal spots (\textit{left panel}) and two non-antipodal spots (\textit{right panel}) corresponding to the data shown in Fig. \ref{fig:data_2spots}.  
}
\label{fig:corner_2spots}
\end{figure*}

From Fig.~\ref{fig:data_errors_sigma} we also see that different assumptions of the model parameters can lead to largely different parameter constraints. 
For example, the constraints are tightest for the model where the seed photons of the Thomson scattering slab have a non-zero polarization (blue dash-dotted curve).
On the other hand, having a larger spot size and having it as a free parameter (orange solid curve) indicates a less constrained NS geometry, which is as expected. 
The models where the observer inclination $i$ is either lower or higher than in the fiducial model (black dashed and dash-dotted curves, respectively) perform as in Fig.~\ref{fig:data_errors_rel}, showing tighter constraints when the PD is higher ($i$ is higher). 
On the other hand, models where the spot co-latitude $\theta$ is altered (red curves) exhibit no significant variation in the parameter limits.  

We now present the parameter constraints when applying the models with two spots (see Table \ref{table:all_models}) in order to inspect how accurately the spot configuration could be constrained using the polarization data. 
The resulting posterior distributions are shown in Fig. \ref{fig:corner_2spots} for one realisation of data and a 200 ks observation time (corresponding to the data shown in Fig. \ref{fig:data_2spots}).
We see that the observer inclination $i$ and primary spot co-latitude $\theta$ are clearly constrained better in the case of antipodal spots, which is expected due to the higher observed PD for the chosen model parameters. 
In the antipodal case, a 95\% credible interval is 57--65\degr\ for $i$ and 13--30\degr\ for $\theta$.  
In the non-antipodal case, a 95\% credible interval is 33--63\degr\ for $i$ and 14--39\degr\ for $\theta$. 
We also obtained constraints for the secondary spot. 
In the antipodal case, we only got a lower limit for $\theta_{2}$, which is around 150\degr\ (the true value being 160\degr). 
The possibility of having a non-pulsating secondary spot exactly at the southern rotational pole can barely be excluded based on these data.  
On the contrary, if $\theta_{2} = 120 \degr$ (as in our non-antipodal case), we get 106--127\degr\ for the 95\% credible interval, and also $\phi_{12}/2\pi$ is constrained between 0.3 and 0.7.  
Thus, the secondary spot is more accurately located in this case, even though the other parameters are less constrained.

\section{Discussion}\label{sec:discussion}

The capability of an accurate determination of the geometrical angles (observer inclination $i$ and spot co-latitude $\theta$) of the NS is based on the swing of the PA with phase when the angles are changed \citep[see][]{VP04,poutanen20}. 
As examined by \citet{SNP18}, having prior information on $i$ and $\theta$  also leads to improved constraints in mass and radius of the NS (even in the case when the opposite solution with switched angles is already excluded).
The assumed constraints for $i$ and $\theta$ from polarization data in \citet{SNP18}, which were about $4-6 \degree$, appear to be in the same order of magnitude as many of the constraints presented in this paper for the longest exposure times. 
Therefore, the accuracy improvement in mass and radius measurements could be the same order of magnitude as found there, which is at least a few per cent level in the measured mass and radius.
However, the true configuration of the spot and the observer strongly affects these estimates as seen in Fig. \ref{fig:data_errors_sigma}. 

There are also a few caveats in the presented error estimates and parameter limits. 
The model for the polarized radiation is based on Comptonization in the Thomson scattering limit for optically thin NS atmospheres \mbox{\citep{VP04}}. 
However, the formalism for Compton scattering in a hot slab should be applied for a more accurate model \mbox{\citep[see e.g.][]{PS96}}. 
The choice of the scattering formalism may affect the predicted measurement accuracy of the Stokes parameters and NS geometry, especially if the modelled PD is different. 
Nevertheless, the effects of having different model parameters have already been accounted for, which shows how the change in PD would affect the results.   
However, we also note that in the most accurate simulations, more of the model parameters (both in the NS and emission models) should be kept free.

When inspecting and visualising the errors in the measured Stokes parameters in Sect. \ref{sec:sim_data}, we calculated the broadband values of $q$ and $u$ using a weighted average. 
We emphasise that while this approach provides a reasonable estimate for the total measurement accuracy it still neglects, for example, the details regarding the energy dispersion of the detector. 
The same also applies for the fitting of the data, where we merged the data into eight different energy channels, instead of applying a full spectro-polarimetric forward-folding fit to $Q$ and $U$ (the latter should be used to properly handle the energy dispersion).  
However, this also implies that our fits are not very prone to the bias caused by small bins having the minimum detectable polarization (MDP, see \citealt{WEO10} for definition) much larger than the PD. 
The usual assumption of Gaussian error distribution in Stokes $Q$ and $U$ is not expected to be valid for that case \citep{mikhalev18}. 
In addition, we simply assume that $q$ and $u$ are uncorrelated and do not employ the bivariate normal distribution presented for them in \citet{KCB15}. 
These effects are not expected to largely influence the estimated errors and the parameter constraints, but they should be accounted for when considering actual observations. 
We also caution that the real exposure time can be restricted by the varying length of the AMP outburst, leaving only the shorter exposure times presented here applicable for such sources.

\section{Conclusions}\label{sec:conclusions}

We have simulated the X-ray polarization data for AMPs that can be detected with IXPE. 
In order to describe AMPs spectral and polarization properties simultaneously, we have also presented a revised Thomson scattering model from \citet{VP04} combined with the Comptonization spectral model \textsc{simpl}.   
We found that the broadband relative error in the measured Stokes parameters $q$ and $u$ attains a 20--60\% level when observing a source like \source\ with a 500~ks exposure time with data binned in 19 phase bins. 
However, the accuracy strongly depends on the model parameters, particularly on the relative configuration of the observer and the emitting hotspot. 

We also determined the constraints for the NS parameters when fitting the simulated polarization data. 
We conclude that NS geometry, particularly the observer inclination and the hotspot co-latitude, could be determined with better than $10\degr$ accuracy for most of the models considered when observing at least with a 500~ks duration. 
This also indicates further constraints on the NS mass and radius, and thus on EOS, when modelling the X-ray pulse profiles. 
In addition, we found that polarization data could also be used to constrain the location of the secondary spot and probe the magnetic field geometry if the secondary spot is not obscured by an accretion disc.  
This could further improve the mass and radius constraints. 
The methods presented here can also be straightforwardly applied when analysing the soon upcoming observations by IXPE. 
On the other hand, the presented errors and parameter limits can be used when inferring the observational strategy of IXPE or other X-ray polarization missions such as eXTP.

\begin{acknowledgements}
This research was supported by the Academy of Finland grant 333112, the Doctoral Programme in Physical and Chemical Sciences of the University of Turku (TS), the Magnus Ehrnrooth Foundation (KK), and the Italian Space Agency (ASI) through the agreement ASI-INFN-2017.13-H0 (LB). 
VL, SST, and JP acknowledge support from the Russian Science Foundation grant 20-12-00364 for the development of the theoretical model (Sect. 2). 
TS thanks Joonas N\"attil\"a for useful discussions. 
The computer resources of the Finnish IT Center for Science (CSC) and the FGCI project (Finland) are acknowledged.  
\end{acknowledgements}

\bibliographystyle{aa}
\bibliography{allbib}

\end{document}